# Beyond Scores: Explainable Intelligent Assessment Strengthens Pre-service Teachers' Assessment Literacy

Yuang Wei
Institute of Artificial Intelligence for Education
East China Normal University
Shanghai, China
Department of Computer Science
National University of Singapore
Singapore, Singapore
philrain@stu.ecnu.edu.cn

Fei Wang
Department of Computer Science
National University of Singapore
Singapore, Singapore
wang-fei@nus.edu.sg

Yifan Zhang
Department of Computer Science
National University of Singapore
Singapore, Singapore
yifan.zhang_@u.nus.edu

Brian Y. Lim
Department of Computer Science
National University of Singapore
Singapore, Singapore
brianlim@nus.edu.sg

Bo Jiang*
Institute of Artificial Intelligence for Education
East China Normal University
Shanghai, China
bjiang@deit.ecnu.edu.cn

## Abstract
Assessment literacy (AL) is essential for personalized education, yet difficult to cultivate in pre-service teachers. Conventional teacher preparation programs focus on theoretical knowledge, while digital assessment tools commonly provide opaque scores or parameters. These limitations hinder reflection and transfer, leaving AL underdeveloped. We propose XIA, an eXplainable Intelligent Assessment platform that extends statistics-informed support with visualized cognitive diagnostic reasoning, including contrastive and counterfactual explanations. In a pre-post controlled study with 21 pre-service teachers, we combined quantitative tasks and questionnaires with qualitative interviews. The findings offer preliminary evidence that XIA supported reflection, self-regulation, and assessment awareness, and helped reduce assessment errors. Interviews further showed a shift from score-based judgments toward evidence-based reasoning. This work contributes insights into the design of intelligent assessment tools, showing how explanatory scaffolding can bridge assessment theory and classroom practice and support the cultivation of AL in teacher education.

*Corresponding author.

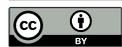

## 1 Introduction

As education increasingly shifts toward personalization and data-driven practices, a central challenge lies in how teachers can effectively interpret and act upon assessment data [7, 57, 89, 95, 125, 136]. Emerging assessment technologies, such as computer adaptive testing (CAT) [62, 86] and cognitive diagnostic assessments (CDAs) [78, 83], are capable of revealing fine-grained knowledge mastery and cognitive barriers, offering teachers data-driven evidence for personalized instruction. Yet these potentials often fail to materialize in practice: teachers—particularly pre-service teachers (teacher candidates who have not yet taken on full professional teaching positions)—struggle to translate outputs filled with statistical indices and model parameters into actionable instructional insights [114, 117]. This difficulty stems not only from the technical nature of outputs but also from the absence of scaffolds that support teachers in transforming outputs into teaching decisions. To unlock the value of diagnostic data, teachers must move beyond reading scores, engage in reflection to build a deeper understanding of student assessment and how it derives learning, and adjust teaching strategies accordingly. [108, 109, 131, 148].

Against this backdrop, there is an urgent need to strengthen teachers' assessment literacy (AL) [29, 39, 46, 108, 119, 128]. Drawing on existing conceptual frameworks, AL can be conceptualized as a multidimensional capability that combines assessment-related knowledge, interpretive and evaluative skills, and cognitive and reflective dispositions [12, 54, 148]. AL goes beyond technical mastery of tools, as it concerns whether teachers can interpret data, engage in reflection, and make instructional decisions in authentic classroom contexts [148]. It also carries a developmental and temporal dimension, resonating with the notion of sustainable assessment: teachers are expected to sustain this capability over time and use it to guide successive cycles of instructional action [5, 12]. Yet current approaches to cultivating AL often remain overly theoretical, emphasizing knowledge transmission while neglecting cycles of practice, feedback, and reflection in authentic tasks [23, 33, 70]. In addition to courses and training programs, reflective practice is a pathway closely aligned with the essence of AL, enabling teachers to refine their judgment and awareness [30, 116, 148]. Effective reflection, however, rarely occurs in a vacuum, as it requires data evidence or contextual triggers to avoid lapsing into intuition and anecdote [13, 96]. CDAs and their underlying cognitive diagnostic models (CDMs) provide such opportunities: they externalize what students know, where they struggle, and so on, turning these into interpretable evidence for teacher reasoning and reflection [36, 78]. In principle, they can serve as powerful carriers for cultivating AL. In practice, however, existing CDA tools often fall short: their inferential processes are opaque, outputs overly technical, and results presented as fixed endpoints rather than as starting points for teacher reflection and instructional design [87, 143]. As a result, pre-service teachers frequently revert to intuition or score-based judgments, hindering the systematic development of assessment awareness and practice.

Lack of explainability is an essential cause of these limitations—teachers see conclusions of CDA tools but not the reasoning pathways or actionable "what-if"s. In practice, these model outputs are often surfaced through teacher dashboards [77, 97] or learning analytics interfaces [44, 133] that make process statistics and model summaries more accessible. However, such systems seldom externalize the evidence chain that connects item-level responses to latent skills and rarely provide actionable explanations that allow teachers to rehearse instructional decisions. In parallel, work on explainable AI (XAI) in education has emphasized algorithmic transparency or student-facing feedback [71]. Comparatively less research targets teacher-facing learning and diagnostic reasoning, where explanations must align with instructional workflows rather than merely reveal model internals [56]. We argue for an explanation design that directly addresses teachers' cognitive needs: responding to questions such as, "Why did the model produce this diagnosis rather than another?" and "Based on my own understanding, what should this student's performance look like?" Contrastive and counterfactual explanations can address these questions by revealing key evidence and enabling teachers to situate results within their own pedagogical reasoning. Such explanations are not meant to replace teacher decision-making, but rather to scaffold reflection, recalibrate assessment awareness, and support instructional exploration [43, 148]. While XAI has shown promise in domains such as medicine and finance, how explanatory mechanisms can be embedded into teacher learning and reflection remains underexplored in education [19, 40, 129].

To address these challenges, we conducted formative interviews with in-service teachers and reviewed existing literature [28, 38, 43, 54, 148], from which we distilled two design requirements: (R1) align decision-support with classroom tasks and (R2) make the evidence chain underlying diagnostic results visible, and articulated three explanatory design principles for teacher-facing CDAs: clarity and traceability, sufficiency with parsimony, and actionability. Guided by these requirements, we designed and implemented XIA (e**X**plainable **I**ntelligent **A**ssessment), a platform that uses explanatory scaffolding to foster AL in pre-service teachers. XIA extends beyond conventional statistics-informed support by visualizing diagnostic reasoning and incorporating contrastive and counterfactual explanations, helping teachers not only see what the model concludes but also understand why, what if conditions changed, and how results differ from their own intuition.

We evaluated XIA in a single-session pre-post controlled study with 21 pre-service teachers, combining quantitative tasks, questionnaires, and semi-structured interviews. The results provide initial evidence that XIA fostered growth in reflection, self-regulation, and assessment awareness, and was associated with more accurate judgments and smaller assessment errors. Interview analyses further indicated a shift from score-oriented judgments toward evidence-based explanatory reasoning.

The main contributions of this work are threefold:

- **Design knowledge for teacher-facing explainable assessment tools.** Drawing on formative interviews with in-service teachers and prior assessment literature, we produce two validated design requirements and three explanatory design principles for designing teacher-facing explainable CDA interfaces. These constitute actionable design knowledge for embedding explainability into assessment systems that support diagnostic reasoning.
- **A system architecture for explainable assessment support.** We contribute a system-level solution that operationalizes these principles within an intelligent teacher assessment platform. The platform provides a reference architecture—linking CDA-based learner modeling, explanation generation, and teacher-centric interaction design—that other assessment tools can adopt. Preliminary usability and learning-effectiveness evidence further validates the feasibility of this architecture.
- **Empirical evidence linking explanatory scaffolding to components of assessment literacy.** Through a controlled pre-post study with 21 pre-service teachers, we provide preliminary evidence that explanatory scaffolding improves specific sub-components of AL. These findings contribute empirical knowledge about how explanation modalities shape teacher diagnostic reasoning and awareness.

## 2 Related Works

As educational paradigms have shifted from standardization and uniformity toward precision and personalization, large-scale personalized assessment has become an important trend in educational research and practice [9, 18, 32]. Teachers serve as the key implementers of assessment, and their AL is essential for the effective realization of personalized instruction [39, 148]. However, pre-service teachers often lack the ability to interpret assessment data and to translate it into instructional decisions and reflective practices, which has become a major bottleneck in the implementation of personalized learning [39, 110, 137]. To clarify the positioning and contribution of this study, this section reviews three areas: the background and challenges of personalized assessment, the importance and difficulties of cultivating teachers' AL, and the potential as well as current limitations of CDA in supporting teachers' professional growth.

## 2.1 Personalized Assessment

Personalized learning has become a central direction of educational reform, aiming to accommodate individual differences in knowledge, skill, and cognition through technology-enhanced approaches [10, 52, 120]. In parallel, personalized assessment at scale has emerged to deliver precise diagnostics [90, 128] that support formative assessment and instructional decision-making.

This vision is partially realized in large-scale exams and commercial platforms. Testing systems like GRE [50], PISA [107], and TOEFL [130] use CAT based on item response theory (IRT) to enhance efficiency and precision. Platforms such as Khan Academy [3], Knewton [111], and Duolingo [27, 92] adopt multidimensional IRT and short-cycle diagnostics to monitor progress.

In research, three major trajectories emerge. First, diagnostic models like CDMs [87] and knowledge tracing (KT) [1] infer fine-grained mastery and barriers and recent work integrates CDMs with deep learning for improved adaptivity [143, 144]. Second, natural language processing and generative models enable automation and personalization of assessment. Applications include automated essay scoring [82], open-ended response analysis [98], and dialogic feedback via large language models (e.g., GPT [63], BERT [140, 142]), with uses across fields like language and medical education [4]. Third, explainable, teacher-oriented systems aim to embed assessment into practice using dashboards [66], visualizations [55], and interactive prompts [106] to support interpretation and instructional decisions.

Despite these advances, three challenges remain. First, the scale and complexity of assessment data exceed many teachers' interpretive capacity, especially pre-service teachers with limited classroom experience [147, 148]. Second, outputs are often technical or statistical, lacking pedagogical meaning and limiting actionability [94, 104]. Third, most systems prioritize learner-facing design while overlooking teacher-facing supports such as explanatory interfaces or reflective scaffolds [80, 105]. This disconnect diminishes their classroom impact. Consequently, advancing personalized assessment requires tools that combine precise diagnostic modeling with explainability and pedagogical usability. Enabling pre-service teachers in particular to meaningfully understand and act upon assessment results is critical for bridging the "assessment-reflection-instruction" cycle that underpins effective personalized learning.

## 2.2 Assessment Literacy

Teacher AL refers to the knowledge, skills, and dispositions that enable teachers to interpret student learning, make instructional decisions, and reflect to improve outcomes [54, 124, 148]. It is not a static competency set but a dynamic process involving goal-setting, tool selection, data interpretation, and instructional adaptation. AL is widely seen as a cornerstone of teaching quality [2, 108, 123].

Developing AL goes beyond technical knowledge, relying on deeper cognition and awareness [31, 88, 115, 146]. Xu and Brown [148] argue that teacher learning primarily occurs through reflective practice and participation in professional communities. In this view, reflection includes both retrospective analysis and prospective planning. It closely links with self-regulation, as teachers identify problems, revise strategies, and refine practice. Also crucial is assessment awareness—the ability to enact policy principles and balance competing demands. Herppich et al. [54] extend this by proposing a structural model that incorporates judgments of process-oriented features (e.g., motivation, strategy use, emotional states), highlighting assessment as an ongoing professional process guiding instruction.

Despite its importance, major challenges remain. In-service teachers often rely on intuition and scores, with limited training in modern data use [42]. Pre-service teachers face sharper obstacles: lacking classroom experience, they struggle to apply theory in practice [31, 51, 109]. Though most programs introduce basic assessment concepts, few offer sustained opportunities for practice, resulting in weak awareness and underdeveloped competence [23, 51]. To address this, scholars advocate shifting from static competency models to dynamic approaches that embed reflection, practice, and feedback in authentic tasks. These have been shown to strengthen assessment—instruction links and foster confidence in data-informed reflection and intervention [23, 75]. Yet, implementation remains limited. Many courses are overly theoretical, and essential supports—like explanatory dashboards, visual tools, and collaborative reflection—are often lacking. Without these, personalized assessment outcomes rarely translate into meaningful instructional change.

## 2.3 Cognitive Diagnostic Assessment Methods

CDA is a class of assessment methods grounded in cognitive psychology. It aims to reveal learners' latent knowledge structures, identify mastery and cognitive barriers within specific dimensions, and support precise instructional intervention [36, 78]. Compared with score-oriented assessments, CDA emphasizes "what students know, what they do not know, and where the obstacles lie," offering fine-grained, multidimensional diagnostic insights [32, 78, 122]. This orientation aligns naturally with the development of teacher AL. Beyond technical proficiency, AL involves understanding goals, selecting tools, interpreting data, and reflecting on instruction based on diagnostic results [45, 48, 148]. In CDA, teachers construct meaning through practice: during design, they may co-define the Q-matrix linking items to knowledge components (KCs); during interpretation, they translate model outputs into actionable judgments. This process builds teachers' data-use capacity [22, 24, 79], systematic assessment awareness [15, 37, 148], and reflective practice across the assessment—instruction cycle [30, 37, 69].

CDMs form the technical foundation of CDA [78, 127], including models like IRT [149], DINA [35], and NeuralCD [144]. These map student responses to a Q-matrix to infer mastery patterns across KCs. CDMs are now widely applied in educational assessment and support personalized learning and formative feedback [143]. Yet, support for teachers remains limited. Inference processes are often opaque, outputs are presented in technical formats misaligned with classroom discourse [21, 61, 81], and many tools treat results as endpoints rather than starting points for reflection or instruction [60, 143]. Thus, while CDMs aid student diagnostics, their role in enhancing teacher AL—especially among pre-service teachers—remains constrained. To address this, CDA tools must move from automation toward explainability. Visual and explanatory designs can help teachers interpret outputs and develop reflective, data-informed instructional decisions.

## 3 Formative Study: Understanding the Information Teachers Need for Student Assessment

### 3.1 Objectives and Participants

AL requires teachers not only to accurately judge students' learning status but also to interpret assessment data and make evidence-based instructional decisions and reflections [124, 148]. However, prior research shows that pre-service teachers often lack these abilities, particularly sensitivity to question characteristics, error patterns, and learning differences [38, 54]. To ensure that our intelligent assessment platform aligns with authentic teaching practices and effectively addresses the needs of AL development, we first conducted interviews to identify the key types of information emphasized by experienced teachers in daily assessments and to explore their attitudes toward intelligent assessment methods.

We interviewed seven in-service teachers in China, with teaching experience ranging from 3 to 32 years. Four were from high schools (including two key schools) and three were from middle schools; five were specialized in mathematics and two in technology. Detailed subject areas and teaching experience are provided in Appendix A Table 2.

### 3.2 Data Collection

The interviews followed a semi-structured format, with questions grounded in core dimensions of teacher AL [108, 148]. Three topics are covered in the questions:

- Q1. Assessment processes without system support: "Based on your experience, if you were given an unknown student's test paper, how would you evaluate the student's mastery of different knowledge components?"
- Q2. Reasoning about errors: "If the test only contained multiple-choice and fill-in-the-blank questions, in what ways would you infer the possible reasons for the student's mistakes?"
- Q3. Attitudes toward intelligent assessment: "If an online platform provided AI-generated assessment information, would you want to know why the system produced such results? Would you question it? Why or why not?"

### 3.3 Data Analysis

We applied inductive qualitative content analysis [41, 59]. Two researchers independently conducted open coding of the verbatim transcripts to identify meaning units related to teachers' assessment reasoning and information needs. After the initial round, the coders compared and discussed their codes to resolve discrepancies. When the two researchers could not reach consensus, disagreements were adjudicated by a senior qualitative researcher before developing a shared codebook. The codebook was then iteratively refined through constant comparison across cases, and related codes were grouped into higher-level categories. The analysis revealed that teachers most frequently attended to question difficulty (7/7), correctness of student answers (7/7), whether questions exceeded the intended scope (6/7), topic alignment (6/7), and typical error patterns (6/7). Representative quotations illustrating these themes include: question difficulty—"*If students get items wrong or a test score is low, I check whether the item or the test overall was too difficult.*" (InS-T 4); correctness of student answers—"*In practice, I judge mastery mainly by the correctness rate on the corresponding items.*" (InS-T 2); whether questions exceeded the intended scope—"*For grade-level tests, I check whether the questions go beyond what they were supposed to learn.*" (InS-T 1); topic alignment—"*In examinations, I check whether the test items align with the content currently being studied; if they go beyond the intended scope, I do not attribute poor performance to the students themselves.*"(InS-T 5); and typical error patterns—"*To judge whether knowledge is firm or there is a misconception, I look for choices/answers that match distractors.*" (InS-T 3).

Regarding attitudes toward intelligent assessment, all teachers (7/7) expressed a desire to understand the reasoning behind AI-generated results, and most (6/7) indicated they would challenge the system when outputs conflicted with their own judgments. These findings underscore the importance of explainability and transparency for teacher adoption. Representative quotations include: desire to understand reasoning—"*If the system says a student has poor mastery of a particular knowledge component, I want to know why—show me the evidence and how it arrived at that conclusion, not just a score.*" (InS-T 6); willingness to challenge conflicting outputs—"*When the tool's result conflicts with my evaluation, I will question it and look for supporting evidence; unless the explanation is convincing, I will rely on my professional judgment.*" (InS-T 4).

### 3.4 Design Requirements

To refine our design requirements, we also drew on prior research. Scholars have long emphasized that effective assessment goes beyond assigning scores: teachers benefit from item-level information—such as difficulty, discrimination, and common error patterns—to inform instructional decisions [14]. Moreover, AL involves the ability to interpret data in relation to individual and class performance, and to align diagnostic results with instructional goals [108, 148]. These insights suggest that assessment tools should ideally not only assist evaluation, but also support teachers in developing assessment-related expertise. Research on AL has also indicated that when diagnostic outputs are presented as opaque end-results, teachers may feel confused. In contrast, when tools clarify the reasoning behind a diagnosis and link it to instructional options, teachers are more likely to benefit [54, 88]. In this sense, clear explanation facilitates the translation of diagnostic information into actionable teaching strategies [8].

Our formative interview analysis revealed similar patterns of need in real teaching contexts, along with two fundamental questions that repeatedly emerged when teachers engaged with diagnostic results: "Why this diagnosis rather than another?" and "What might happen if an intervention were made?" These questions reflect contrastive and counterfactual reasoning, respectively. In real-world assessment, when teachers wish to identify potential sources of error or weigh the relevance of certain findings, they often engage in comparative reasoning to evaluate plausibility or appropriateness—prompting the contrastive question. Likewise, when teachers consider how learning outcomes might change under different instructional actions, they are engaging in counterfactual thinking. These observations led us to consider whether contrastive explanations [17, 134, 141] and counterfactual explanations [132, 139, 141] might help teachers better understand diagnostic outputs and, in doing so, support the development of AL.

Therefore, we aim to provide clear and specific diagnostic information and explanation as scaffolds—not only to assist teachers in making assessments, but also to encourage self-evaluation and reflection, and to strengthen their understanding of how evidence leads to conclusions [99, 121]. At the same time, diagnostic reasoning interfaces must remain understandable and compatible with classroom workflows, especially given that most teachers are unfamiliar with the underlying technical details [71, 135]. These insights informed two overarching design requirements:

R1: **Decision-support information**. *Aim and user*—To enable pre-service teachers to quickly situate a learner's performance and plan next instructional steps using signals beyond a single score. *Context*—Teacher-facing, formative assessment with limited time/technical expertise; item- and class-level data available. *Mechanism*—Aggregate question-level metadata and inferred cues and present them with individual—class comparisons; enable drill-down from an overview to specific items and exemplars. *Rationale*—Surfacing teacher-relevant and comparable signals reduces search effort, situates a case within cohort patterns, and is identified in the assessment literature as a prerequisite for valid interpretation and near-term instructional planning [14, 108, 148].

R2: **Visualized diagnostic reasoning**. *Aim and user*—To enable pre-service teachers to see how evidence supports diagnostic conclusions and to reflect on assessment results. *Context*—Complex or borderline assessment cases where users are unfamiliar with model internals. *Mechanism*—Expose the evidence→interpretation links in a diagnostic-reasoning view (with contrastive and counterfactual exploration) and link each step back to items and artifacts. *Rationale*—Transparent, traceable reasoning helps teachers interpret learning differences and connect diagnostic findings to concrete pedagogical choices, as recommended by research on assessment competence and diagnostic explanation [8, 54, 88].

## 4 XIA System Architecture

### 4.1 System Overview

The XIA platform combines a front-end interface with a back-end computation module. Using student responses and item metadata, it applies statistical analysis and cognitive diagnostic modeling to provide decision-support information and visualized diagnostic reasoning (Fig. 1).

XIA offers two complementary interfaces: **(1) Instructional Decision-Support Interface** presents test, knowledge, and item-level analyses, such as accuracy, difficulty, and knowledge mastery distributions. Addressing **R1**, it highlights assessment signals that help teachers identify whether issues arise from students, instruction, or item design. **(2) Diagnostic Reasoning and Explanation Interface** visualizes the diagnostic process and, following **R2**, generates contrastive and counterfactual explanations. These allow teachers to examine alternative assumptions, contest model outputs, and reflect on student learning.

Together, the two interfaces connect statistical outputs with pedagogical sensemaking, ensuring both actionable diagnostic insights and transparent reasoning support.

### 4.2 Front End

The XIA front-end provides two complementary interfaces (Fig. 2).

*4.2.1 Instructional Decision-Support Interface.* The Instructional Decision-Support Interface presents assessment data that teachers frequently rely on (see section 3.3 and 3.4), including question difficulty, student accuracy, typical error patterns, comparisons between question difficulty and class ability, and differences between individual and group performance (see Table 1). These comparisons help teachers identify whether observed issues originate from students, instructional practices, or the design of the test questions, and support judgments about whether questions may exceed curricular scope or deviate from intended constructs.

In addition to these teacher-identified indicators, the interface also provides basic statistics that offer teachers a broader context for interpreting results. Specifically, it includes an overall summary of student performance and the relative weight of KCs in the test (Test Analysis Overview), comparisons of knowledge mastery across topics (Knowledge Analysis), and indices of item discrimination (Question Analysis). Moreover, the system generates natural language teaching suggestions based on students' specific performance, helping teachers translate diagnostic results into actionable instructional decisions.

*4.2.2 Diagnostic Reasoning and Explanation Interface.* In addition to the instructional decision-support interface, XIA provides a diagnostic reasoning and explanation interface that makes the model's inference process transparent. This interface presents the reasoning path in a visual form and offers two complementary types of explanations. Contrastive explanations address the question "why result P rather than Q?" by comparing outcomes under different response patterns to highlight the evidence that drives the model's judgment. For example (Fig. 2 R2), when a student answers question 1 correctly but questions 2 and 3 incorrectly, the model estimates mastery at 40%; when the correctness of question 1 and question 2 are swapped, the estimate decreases to 34%, revealing that question 1 carries more diagnostic weight than question 2. In contrast, counterfactual explanations address the question "what would the outcome be if the underlying mastery assumptions were different?" by assuming an alternative mastery level and generating the most likely response pattern consistent with that assumption. For instance (Fig. 2 R2), when the model diagnoses a student's mastery as 52% but a teacher believes it should be closer to 34%, the system produces a hypothetical scenario in which, at 34% mastery, the student would likely answer all three questions incorrectly.

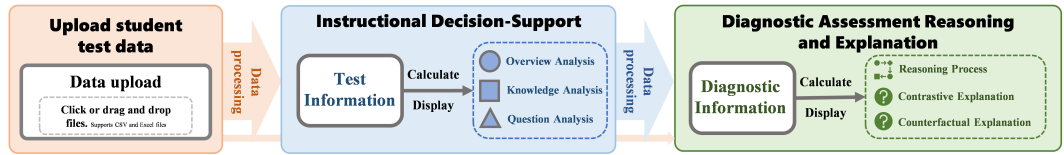

Figure 1: Workflow of the XIA platform. Teachers first upload student test data in CSV or Excel format. The system processes the data and generates teaching decision support information, including overview analysis of test performance, question-level analysis, and knowledge mastery analysis. Based on this, diagnostic assessment reasoning and explanation modules provide detailed diagnostic information, reasoning process visualization, and both contrastive and counterfactual explanations to support teachers' decision-making.

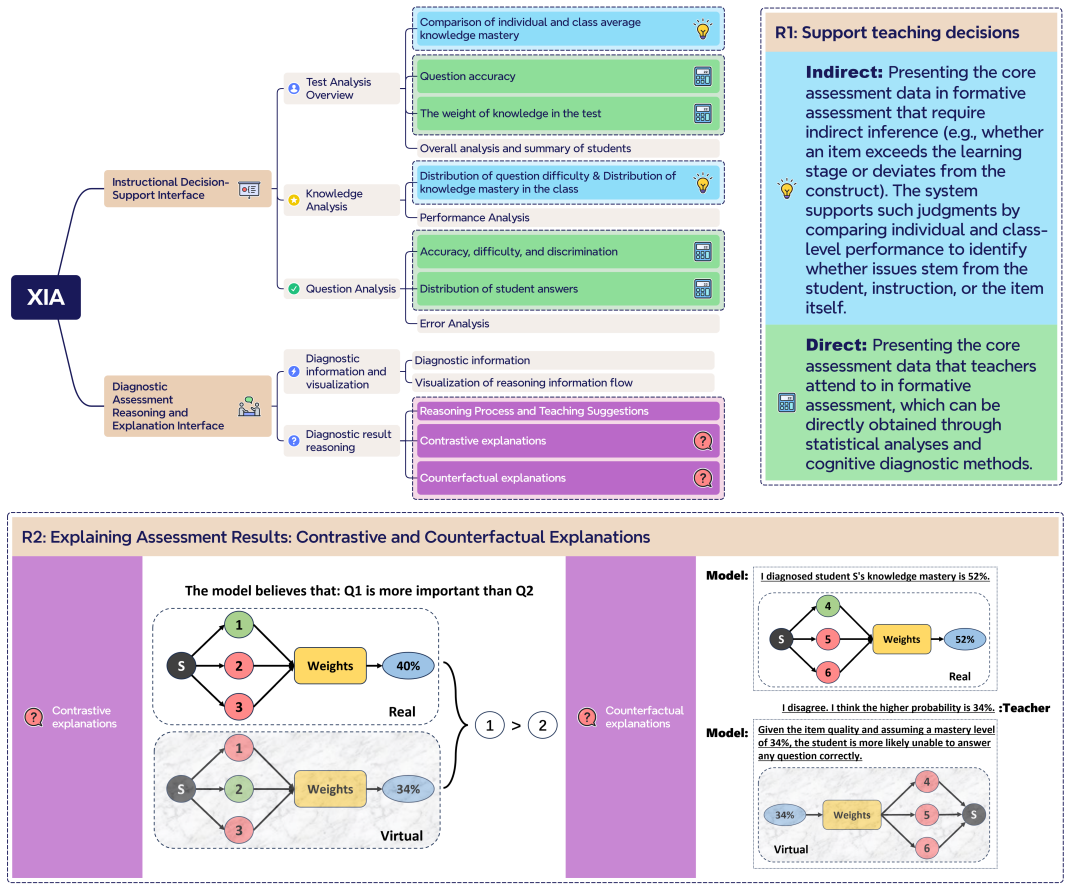

Figure 2: Overview of the XIA Front-end. The system consists of two main components: (1) an interface for supporting instructional decision-making, which provides teachers with core assessment data such as question difficulty, accuracy, class-student comparisons, and error analysis (R1); and (2) an interface for visualizing diagnostic reasoning, which explains model outputs through reasoning flows, contrastive explanations, and counterfactual explanations to foster reflection and instructional adjustment (R2).

The explanatory design of XIA is not about showing more information; it is about enabling teachers to build and calibrate a causal mental model that links observable evidence to latent skills or errors and, ultimately, to a defensible diagnostic conclusion. We operationalize explanation quality with three mutually reinforcing criteria—*clarity and traceability*, *sufficiency with parsimony*, and *actionability*.

Fig. 4 provides an overview of the explanatory interface ( D - H ), which operationalizes our three criteria. *Clarity and traceability* are realized through a double-panel layout. *Sufficiency with*

Table 1: Key indicators in the Instructional Decision-Support Interface

| Indicator | Meaning | Use | Source (Teacher Need) |
| --- | --- | --- | --- |
| Question difficulty | Reflects the relative challenge level of an item. | Used to judge whether items are too easy or too hard. | Teachers attended to **question difficulty** (7/7). |
| Student accuracy | Shows individual students' correctness across items. | Helps identify struggling learners and overall gaps. | Teachers attended to **correctness of student answers** (7/7). |
| Typical error patterns | Reveals frequent misconceptions. | Highlights common misunderstandings from distractor choices. | Teachers attended to **typical error patterns** (6/7). |
| Comparisons between question difficulty and class ability | Examines whether item difficulty aligns with class proficiency. | Used to judge if items are disproportionately harder than the class's overall ability, suggesting they may exceed scope. | Teachers attended to **whether questions exceeded the intended scope** (6/7). |
| Differences between individual and group performance | Distinguishes personal vs. collective performance issues. | Used to identify whether low performance stems from individual students or reflects a systematic class-wide issue (topic alignment). | Teachers attended to **topic alignment** (6/7). |

*parsimony* is addressed by presenting a small, readable subset of evidence by default and deferring nonessential details until the teacher actively requests them. *Actionability* is achieved by allowing teachers to click to flip an item response and inspect the corresponding diagnostic conclusion in that scenario; when the system's diagnosis conflicts with their own judgment, teachers can input their preferred answer to see how and why the resulting conclusion differs. A step-by-step description of these interaction flows is provided in Appendix B.1. We designed this form of explanation, on the one hand, to address the two common questions that emerged from teachers' real-world assessment practice and were summarized in the design requirements in Section 3.4, and on the other hand to help teachers construct a causal mental model that follows an evidence → latent skills or errors → conclusion reasoning path. In this way, we aim to help teachers not only understand why the model produces the current diagnosis, but also explore how results change under different conditions, thereby turning diagnostic outputs from a static end point into a starting point for reflection and dialogue.

### 4.3 Back End

The back-end consists of three modules (Fig. 3): a statistical data processing module, a Neural Cognitive Diagnosis Model (NeuralCD) computation module [144], and an explanation module, which together support the statistical displays and reasoning visualizations in the front-end. The data processing step integrates student response records, question-knowledge mappings (Q-matrix), response matrices, and student/question IDs, which are encoded into one-hot representations as model inputs. The NeuralCD decomposes these inputs into four parameters—knowledge relevancy, student proficiency, question difficulty, and question discrimination—and fuses them through multilayer non-negative fully connected networks. Rather than merely predicting whether a student will answer correctly, NeuralCD serves as a diagnostic engine that estimates students' mastery probabilities across knowledge dimensions and reveals how item characteristics interact with proficiency. These diagnostic outputs form the basis for instructional interpretation. The explanation module then builds on these results by visualizing reasoning chains and generating contrastive and counterfactual explanations, thereby transforming the underlying diagnostic computations into interpretable evidence that scaffolds teachers' reflection and decision-making. The explanation module implements three mechanisms:

(a) **Student representation estimation (posterior embedding update)**: estimating the target student's latent knowledge state from their responses while keeping item parameters fixed;
(b) **Contrastive reasoning**: comparing different response records of the same student to highlight knowledge differences;
(c) **Counterfactual reasoning**: substituting embeddings on specific knowledge dimensions to simulate alternative mastery levels and generate predicted outcomes.

These mechanisms not only power the front-end interactions but also ensure explainability and transparency in the system outputs. Detailed algorithm calculation steps are shown in Appendix B.2.

## 5 User Study with Pre-service Teachers

This study evaluates the effectiveness of the XIA in fostering AL among high school pre-service teachers. We adopted a mixed-methods study, combining quantitative task performance, standardized questionnaires, and semi-structured interviews to examine how XIA supports teachers in judging students' knowledge mastery, promoting instructional reflection, and strengthening assessment awareness. The experiment followed a 3 (group, between) × 2 (time, within) mixed design. The study received approval from the institutional ethics committee, all participants provided informed consent, and all data collection complied with relevant privacy and data protection guidelines.

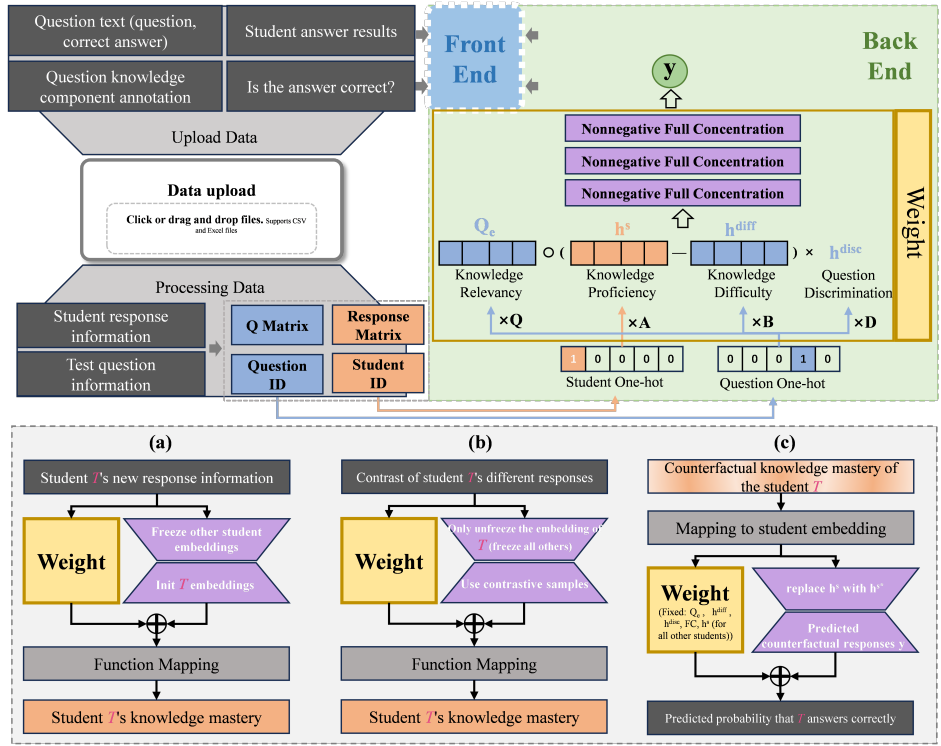

**Figure 3: Back-end architecture of XIA.** The data processing module integrates student responses, item text and correctness, knowledge-component annotations (Q-matrix), and identifiers into structured matrices and one-hot encodings. The NeuralCD decomposes these inputs into four latent parameters—knowledge relevancy ($Q_e$), student proficiency ($h^s$), question difficulty ($h^{diff}$), and question discrimination ($h^{disc}$)—which are fused via nonnegative fully connected layers to estimate the probability of correct response $y$. The explanation module implements three mechanisms: (a) updating the embedding of a new student while freezing others to estimate mastery; (b) contrastive reasoning by generating alternative response subsets to reveal evidential differences; and (c) counterfactual reasoning by replacing embeddings on specific dimensions to simulate alternative mastery levels and recompute predicted outcomes.

## 5.1 Participants and Study Setup

We recruited 21 pre-service teachers in China who had completed coursework in mathematics / technology education and educational foundations, and who had short-term practicum experience but limited classroom assessment practice. The study was conducted online via Tencent Meeting, with surveys administered through Wenjuanxing. Each participant received compensation of $10 per hour. Participants were randomly assigned to one of three groups (see Appendix C.1.1 Table 3 for details.):

- **Control Group (CG, n=7)**: No tool support.
- **Decision-Support Group (DSG, n=7)**: The **experimental group** that accessed the Instructional Decision-Support Interface (Fig. 4 Ⓐ - Ⓒ).
- **Full-Support Group (FSG, n=7)**: The **experimental group** that accessed both the Instructional Decision-Support Interface (Fig. 4 Ⓐ - Ⓒ) and the Diagnostic Assessment Reasoning and Explanation Interface (Fig. 4 Ⓓ - Ⓗ).

To avoid learning effects, pre- and post-tests used parallel test forms covering the same KCs with similar difficulty but different questions, all of which were high school mathematics problems. Four standardized questionnaires were administered in parallel form across both sessions, with only minor contextual wording changes.

## 5.2 Assessment Baselines and Questionnaire Design

*5.2.1 Setting the Baseline for Students' Knowledge Mastery.* Because student knowledge states are latent constructs with no directly observable ground truth, we approximated a quasi-ground-truth by aggregating a broader set of response and question data beyond the experimental tasks. Each student contributed about 25 questions per KC, yielding 3,318 response records in total, collected from a high school in China. Prior work suggests that such test length and coverage are sufficient for reliable CDM estimation [118, 144]. Using this extended dataset, we estimated mastery probabilities with NeuralCDM. To assess the plausibility of this diagnostic reference, we compared the estimated mastery probabilities with students' observable learning outcomes—specifically their KC-level correctness rates. The two sets of indicators exhibited consistent trends: KCs with higher estimated mastery tended to show higher empirical correctness, whereas lower estimated mastery corresponded to more frequent or systematic errors (see Appendix C.1.2). While not constituting a definitive validation of latent abilities, these alignment patterns offer supportive behavioral evidence that the diagnostic reference coheres with students' actual performance structure. The resulting baseline provided a more reliable reference against which we paired pre-service teachers' judgments to evaluate their accuracy and identify sources of discrepancy.

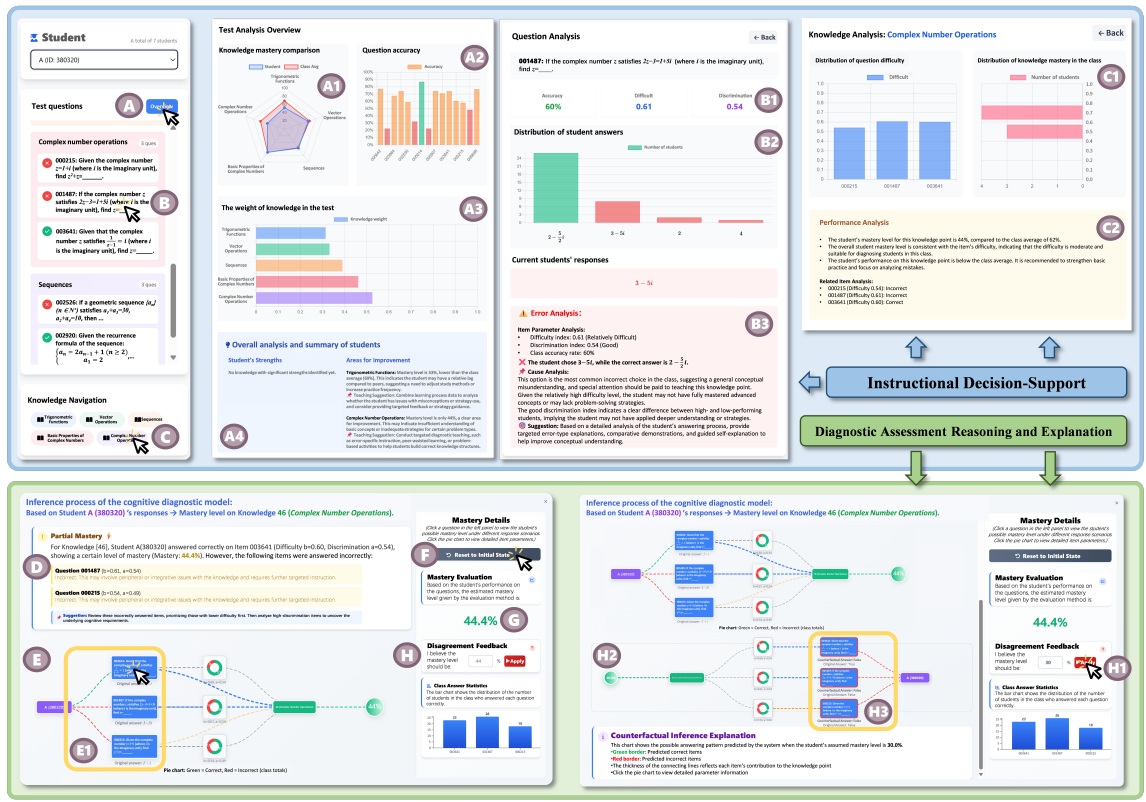

Figure 4: Two parallel functions of the XIA platform interface. Ⓐ - Ⓒ show instructional decision-support views. They provide multi-level statistics, from per-student test summaries to per-knowledge and per-question results, and allow navigation and filtering by student or item through the left panel to support classroom planning and intervention. Ⓓ - Ⓗ show cognitive diagnostic reasoning and explanations. These views visualize the inferred mastery profile and item—KC relations, and present inference chains together with contrastive and counterfactual explanations, helping teachers understand AI-generated diagnoses and reflect on alternative instructional options. See Appendix B.1 for a detailed introduction to each component.

*5.2.2 Questionnaire Design.* To measure changes in multiple dimensions of AL, we adapted four validated instruments from prior work [16, 69, 93, 138]. The instrument had isomorphic pre/post forms and comprised four parts. (1) A scenario-based instructional judgment test adapted from the ALI [93]: we reduced the original pool of items to four multiple-choice scenarios and rewrote them into secondary-mathematics contexts (e.g., diagnosing error causes, judging item discrimination, checking coverage/goal alignment, and comparing individual vs. class performance), preserving the ALI's format but contextualizing it for pre-service teachers; responses were scored by accuracy. (2) A reflective-thinking section based on Kember's four dimensions (habitual action, understanding, reflection, critical reflection) [69]: items were retained in structure but reworded from general study to assessment-design contexts, answered on a 7-point scale (1—7). (3) A teacher SRL section adapted from the Teacher SRL scale [138]: items covering planning, monitoring, time management, task perception, adjustment, reflection, and attribution were rewritten so that statements referred to assessment design rather than generic learning, answered on a 7-point scale (1—7). (4) A conceptions-of-assessment section derived from the CoA-III [16]: the original items were shortened to 13 representative statements capturing alignment, improvement, learning promotion, diagnostic value, accountability, external challenges, and drawbacks; culture- or policy-specific wording was removed while retaining the four core factors, answered on a 7-point scale (1—7). All adaptations followed the original factor structures to maintain construct validity and were reviewed by two experts to ensure contextual appropriateness for pre-service teachers. Furthermore, in the present sample we re-evaluated internal consistency and item discrimination; Cronbach's $\alpha$ ranged from 0.573 to 0.720 and the corresponding ordinal $\alpha$ (Spearman) from 0.661 to 0.814, with the lowest subscale showing $\alpha = 0.573$ but an ordinal $\alpha = 0.712$, indicating overall acceptable reliability of the adapted scales. Full item lists and adaptation notes are provided in Appendix C.2.

*5.2.3 Interview Design.* To complement the quantitative measures, we conducted semi-structured interviews after the post-test to probe participants' information adoption strategies, evaluation rationales, and tool-use experiences. The protocol was informed by observed behaviors and questionnaire results, and adapted across groups to match their experimental conditions. Interviews used open-ended questions with probes to elicit examples of reasoning and reflection, and organized around eight themes: (1) information adoption and trust (original vs. newly introduced), (2) data interpretation and threshold judgments, (3) decision tendencies and adoption strategies, (4) feature use and metacognition, (5) cognitive conflict and reflection, (6) expectations for human-AI interaction, (7) baseline comparison and perceived influence, and (8) usage context and adoption boundaries.

Interview focus varied across groups:

- **Control group**: Items related to system features were rephrased to fit the no-tool context (Themes (1), (2), (3), (5), (8)).
- **Decision-support group**: Five themes related to statistical data (Themes (1), (2), (3), (7), (8)).
- **Full-support group**: All eight themes.

Interviews followed an open-ended question-and-probe format to encourage participants to illustrate their reasoning processes and behavioral changes. Examples of interview questions are shown in Appendix D.2 Table 10.

### 5.3 Study Procedure

The study was conducted online and recorded for later verification. Participants were randomly assigned to one of the three groups and completed the experiment in three phases (Fig. 5). In the pre-test phase, participants independently judged students' knowledge mastery based on their responses and completed baseline questionnaires. During the tool explanation and experience phase, participants in all groups received the same primer on the principles of IRT (a classic approach in CDA) to help them understand how diagnostic results are generated. The experimenter then provided an identical, detailed walkthrough of the system's interactive features and functions for all groups, after which participants were given time to freely explore the platform. During the demonstration and exploration, all groups used the same full XIA interface; the screen displayed the diagnostic results for the two students from the pre-test phase so that everyone could learn the navigation and output semantics without exposure to post-test materials. Throughout the free-exploration period, participants in all groups could ask questions about interface mechanics and result interpretation and received on-the-spot clarification from the experimenter. In the post-test phase, participants evaluated new students on parallel tasks, completed follow-up questionnaires, and joined semi-structured interviews.

### 5.4 Hypotheses

We propose the following alternative hypotheses, informed by prior work on AL and XAI in education [38, 54, 76, 148]. These hypotheses target both participants' performance in assessment tasks and their development in reflective thinking, self-regulated learning, and assessment awareness.

**H1.** XIA improves assessment performance. Specifically, participants will (H1a) show significant reductions in error of their assessments of students' knowledge mastery from pre- to post-test, and (H1b) the experimental groups will outperform the control group in these performance gains.

**H2.** XIA fosters the development of AL. Specifically, participants using the tool will exhibit significant improvements in (H2a) reflection, (H2b) self-regulated learning, and (H2c) assessment awareness.

**H3.** The reasoning and explanation interface further enhances AL compared to decision-support alone. Specifically, participants with access to explanatory reasoning features will demonstrate greater improvements in (H3a) reflection, (H3b) self-regulation, and (H3c) assessment awareness.

**H4.** The experimental groups outperform the control group in AL. Specifically, compared to participants without the tool, those using XIA will achieve greater improvements in (H4a) reflection (H4b) self-regulated learning, and (H4c) assessment awareness.

## 6 Results

We present results on teachers' questionnaire outcomes, assessment accuracy (pre- vs. post-test), and interview findings.

### 6.1 Assessment Literacy Questionnaire Results

All participants successfully completed the ALI-based situational test (AL Inventor [93]), confirming that they shared a baseline level of assessment knowledge. We then compared pre-post changes in three non-knowledge dimensions of AL: reflection, self-regulation, and assessment awareness (Fig. 6).

Since the questionnaire scores were approximately normally distributed, we applied paired-sample t-tests within groups and Welch t-tests between groups.

**Reflection:** As shown in Fig. 6a, both the decision-support group (Mean$\Delta$ = 0.70, $p < 0.05$) and the full-support group (Mean$\Delta$ = 0.86, $p < 0.001$) showed significant pre-post improvements, while the control group showed no change (Mean$\Delta$ = 0.04, n.s.), supporting H2a. Between-group comparisons confirmed that full-support group's gain was significantly larger than control group's ($p < 0.01$), and the decision-support group also outperformed the control group at a marginally significant level ($p = 0.07$). No significant difference was found between the decision-support group and full-support group on reflection, indicating that H3a was not supported. However, full-support group did outperform the control group, providing support for H4a.

**Self-regulation:** Both the decision-support group (Mean$\Delta$ = 0.81, $p < 0.01$) and the full-support group (Mean$\Delta$ = 0.93, $p < 0.01$) improved significantly, whereas control group showed no improvement (Mean$\Delta$ = 0.13, n.s.) illustrated in Fig. 6b, supporting H2b. Between-group tests confirmed that both the decision-support group and full-support group significantly outperformed the control group ($p < 0.05$), but no difference was found between the decision-support group and full-support group. These results support H4b but do not support H3b.

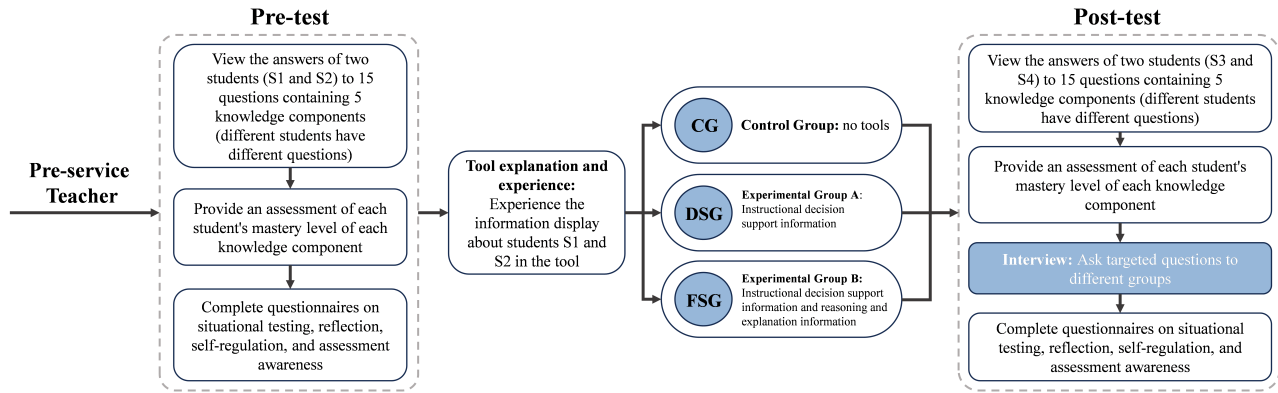

Figure 5: Study procedure and Group Assignment.

**Assessment awareness:** Full-support group exhibited a significant improvement (Mean∆ = 0.68, $p < 0.01$), while the decision-support group (Mean∆ = 0.34) and the control group (Mean∆ = 0.25) did not, see Fig. 6c, supporting H2c. Between-group tests showed that the full-support group's gain was significantly higher than both the decision-support group and the control group ($p < 0.05$), supporting H3c and H4c.

Overall, in this single-session study, these results provide preliminary evidence that XIA can support improvements in reflective thinking, self-regulation, and assessment awareness, with the full-support version showing the most consistent pattern of gains. Given the short duration of the intervention and the small sample size, these findings should be interpreted cautiously as preliminary evidence of the XIA's potential impact. Detailed analyses are provided in Appendix C.1.3 (Tables 7 and 8).

### 6.2 Results on Assessment Accuracy

All participants completed both pre- and post-test evaluations, allowing us to compute mean absolute error (MAE) and root mean square error (RMSE) for each group (Fig. 7). To provide a fair basis for interpreting change, baseline comparability across groups was confirmed: pre-test errors did not differ among groups, with pairwise t- and Mann-Whitney U tests showing no significant differences for either MAE or RMSE (all $p \geq 0.165$).

For the decision-support group, average MAE decreased by 0.033 and RMSE by 0.034 from pre- to post-test; however, these changes were not statistically significant ($p = 0.45$ and $p = 0.47$; Table 5). For full-support group, reductions were substantially larger. Average MAE decreased by 0.063 ($p = 0.009$) and RMSE by 0.067 ($p = 0.023$), with Wilcoxon tests reaching the same conclusions ($p = 0.016$ and $p = 0.047$), indicating that participants made significantly fewer and less extreme errors after using the tool. For control group, average MAE decreased by 0.013 and RMSE by 0.022, with no significant differences ($p = 0.55$ and $p = 0.33$). Some participants even showed increased error in the post-test, leading to minimal overall gains.

Comparing across groups, the pattern indicates that full-support group benefited the most from tool use, followed by modest, non-significant improvements in the decision-support group, while the control group showed little change. Importantly, between-group contrasts of gain scores did not reach significance (Table 6; e.g., full-support group vs. control group: MAE $p = 0.46$, RMSE $p = 0.32$); full statistics are reported in the Appendix C.1.3 Table 5 and 6. In conclusion, the results in this single-session study provide preliminary evidence that reasoning/explanation features support improvements in teachers' assessment accuracy (H1a) and are directionally consistent with H1b, although without statistical significance yet. Given the short duration and the small sample size, the findings need to be considered preliminary but informative for future research.

### 6.3 Interview Results

To analyze the interview data, we applied the same inductive qualitative content analysis approach as in the formative study. Two researchers independently coded the transcripts, compared and discussed their codes to resolve discrepancies, and, when they could not reach consensus, remaining disagreements were adjudicated by a senior qualitative researcher before iteratively refining a shared codebook through constant comparison across cases. This process allowed us to consolidate recurrent patterns into higher-level themes, ensuring methodological consistency and analytic rigor across both studies.

The interview analysis provides preliminary qualitative evidence that aligns with the quantitative patterns, indicating potential ways in which participants engaged with XIA's support. Across conditions, teachers reported moving beyond score-only judgments toward integrating multiple sources of evidence. For instance, one full-support group participant noted, *"I no longer just look at correctness rates-I combine question difficulty and discrimination."* (FSG-01), while another emphasized, *"The numerical values provided by the tool helped me organize my thinking and avoid wavering."* (FSG-03). These reflections suggest that participants began to anchor their evaluations in multiple diagnostic cues rather than relying solely on intuition.

Full-support group participants consistently described the platform as scaffolding their judgments, highlighting calibration and evidence weighting as central practices. For example, one reported, *"I used to rely on my own experience, now I calibrate with the platform information."* (FSG-04). Others referenced combining difficulty, discrimination, and observed accuracy to reach more stable judgments. Such discourse indicates a shift toward routinized, evidence-based reasoning. This qualitative pattern reinforces the group's stronger quantitative gains in reflection and accuracy, pointing to explanatory features as a driver of systematic diagnostic practice.

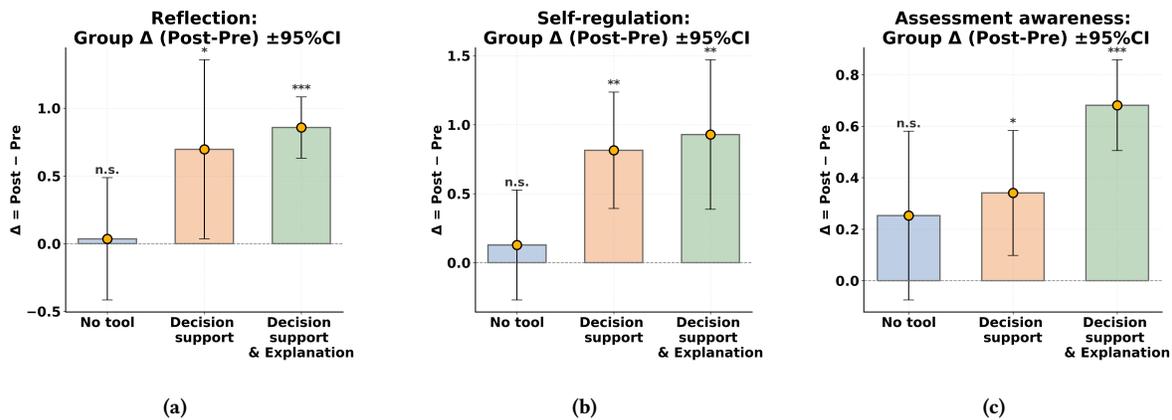

Figure 6: Questionnaire outcomes on three subscales of AL. Bars represent mean gain scores (Δ =Post−Pre) for each group, with error bars showing 95% confidence intervals. Asterisks indicate significance levels: * $p < 0.05$, ** $p < 0.01$, *** $p < 0.001$; n.s. = not significant.

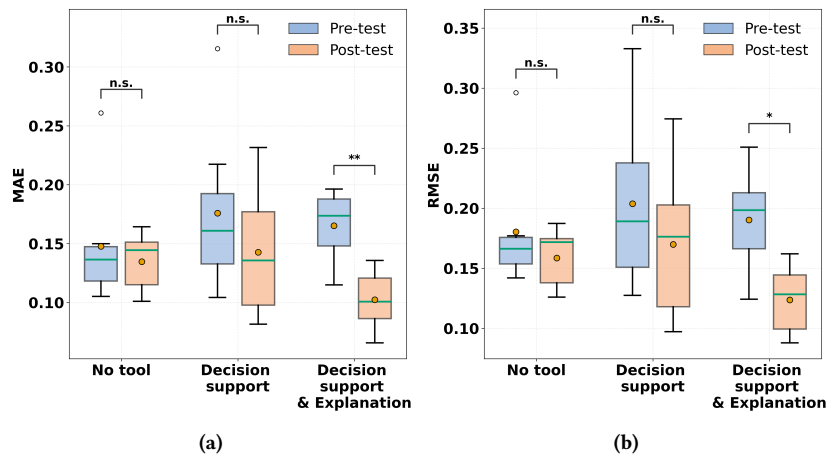

Figure 7: Assessment accuracy outcomes by group and phase. Boxplots show pre- and post-test distributions of mean absolute error (MAE) and root mean square error (RMSE). Both experimental groups (decision-support group and full-support group) showed reductions from pre- to post-test, with the full-support group achieving the largest overall improvement compared to the control group. Asterisks indicate significance levels: * $p < 0.05$, ** $p < 0.01$, *** $p < 0.001$; n.s. = not significant.

Decision-support group participants acknowledged the usefulness of multiple indicators and visualizations, yet their strategies remained less systematic. Their responses often emphasized noticing additional signals rather than integrating them into stable routines. For instance, one teacher remarked, *"Data visualization helps me examine student learning."* (DSG-07), while another described *"switching from single-variable to multi-variable assessments."* (DSG-01). These accounts demonstrate cognitive expansion—recognizing more dimensions of evidence—but with less consistent operationalization. This aligns with decision-support group's more modest quantitative improvements, suggesting that awareness of new indicators is a precursor to deeper diagnostic reasoning.

Control group participants recognized the limits of score-only information and articulated principle-level needs. For example, one noted, *"We need to include process data and student self-reports for a multi-faceted understanding."* (CG-02). Yet, their reflections generally stopped short of specifying operational strategies for judgment, such as how to weigh difficulty or when to recalibrate item quality. This abstract awareness corresponds with the group's limited quantitative change, underscoring the challenge of translating recognition into practice without targeted scaffolding.

Taken together, these interviews offer preliminary qualitative support, triangulating the quantitative results by showing different depths of engagement with the tool's outputs: The control group expressed abstract awareness, the decision-support group expanded their attention but with emerging routines, and the full-support group articulated systematic calibration strategies. This gradient mirrors the quantitative outcomes and reinforces the conclusion that explanatory features in XIA foster deeper reflection, self-regulation, and evidence-based assessment practices. More examples can be found in Appendix D.1.

## 7 Discussion

### 7.1 Effectiveness of the XIA in Developing Pre-service Teachers' Assessment Literacy

Beyond providing preliminary evidence of overall gains, our findings also lend some support to prior claims that task-embedded assessment practice can enhance pre-service teachers' AL [11, 70]. Compared with traditional theory-focused learning, XIA placed pre-service teachers in authentic diagnostic scenarios, where they repeatedly judged student responses and actively compared and interpreted evidence. This kind of learning-in-task appears to draw teachers' attention to key assessment cues [54], such as item difficulty, discrimination, and knowledge coverage. Interview data likewise show that, while working with the system, participants more frequently attended to these multidimensional indicators of assessment validity, rather than relying on intuition or a single score. This pattern aligns with earlier work suggesting that authentic practice is particularly conducive to reflection and transfer [148].

At the same time, the three AL dimensions we examined displayed different patterns of change. Reflection and self-regulation improved significantly in both experimental groups, which is plausible given that these abilities depend on ongoing monitoring, comparison, and adjustment during task engagement [99]. In our study, the task structure required participants to repeatedly review their judgments and modify their strategies; this may help explain why even exposure to statistical information alone was enough to trigger reflective and self-regulatory processes. By contrast, assessment awareness reflects a deeper layer of belief about the purposes, standards, and instructional roles of assessment, and such beliefs typically change only when supported by richer informational environments and more substantial evidence [15]. Consistent with this view, only the full-support group—who received more extensive information—showed significant gains in assessment awareness, while the other groups changed little. These findings tentatively suggest that different layers of AL may be supported by different types of learning experiences: more surface-level, strategic components can be strengthened through task participation, whereas deeper assessment awareness require stronger contextual and informational scaffolding. With respect to assessment accuracy, our results showed only limited within-group improvement and no significant differences between groups. This is consistent with prior work suggesting that accuracy develops as a "slow variable" through long-term practice and repeated calibration [49, 102]. In contrast to abilities that can be activated in the short term, accuracy depends more heavily on exposure to a large number of diverse responses and on gradual sensitization to error patterns. Short-term interventions are therefore more likely to compress extreme misjudgments than to rapidly restructure teachers' overall judgment strategies. This pattern implies that, if the goal is to meaningfully improve assessment accuracy, tools like XIA need to be embedded in sustained, cross-task use that supports ongoing calibration, rather than being confined to one-off or short-cycle training activities.

A further observation is that, even without access to the explanatory interface, the decision-support group still improved in reflection and self-regulation. This suggests that transparency in statistical information itself can contribute to AL. Interview data indicate that many participants reported "shifting from looking only at accuracy to also considering difficulty and discrimination" (e.g., DSG-01, DSG-03) and began to use multiple indicators to check their judgments. In other words, even without deeper reasoning support, rich and structured data can already help pre-service teachers loosen a purely score-based stance and move toward more evidence-based assessment habits. This offers a design implication for dashboard-like tools: when appropriately structured and embedded in task-embedded learning activities, statistical information alone may support certain core components of AL, even if it is not sufficient to reshape deeper assessment beliefs.

### 7.2 Unique Contributions of Explainable Information to Teacher Improvement

Compared with "black-box" diagnostic tools that provide only result-level outputs [6, 43, 47], the explanatory scaffolding in our tool helped make model reasoning more transparent and actionable. Our experimental results provide preliminary evidence that such scaffolding can not only increase teachers' acceptance of the system, but also begin to reshape their reasoning strategies in ways that are visible across multiple levels of evidence.

**Observed performance outcomes:** Performance distributions provide the clearest illustration of this effect. From pre- to post-test, the full-support group showed the strongest downward shift in both MAE and RMSE: median errors decreased, interquartile ranges contracted, and RMSE dropped most sharply. Given RMSE's sensitivity to large deviations, this pattern suggests that participants in the full-support group were not only closer to the correct answers on average, but were also systematically avoiding "outlier" mistakes. The decision-support group also showed a decline in median error, but with less contraction in variance, indicating moderate improvement without consistent suppression of severe misjudgments. The control group displayed virtually no stable change and even showed isolated extreme errors at post-test, suggesting that unguided experience alone was insufficient to build robust assessment competence. Although between-group contrasts on MAE and RMSE did not reach statistical significance (Appendix C.1.3), the consistent distributional shifts and sizeable within-group effects in the full-support group point to the practical importance of the explanatory condition. Moreover, these performance trends parallel the questionnaire results: the full-support group achieved larger gains in reflection and self-regulation, consistent with deeper metacognitive engagement and strategy adjustment during the tasks [91, 113].

In other words, participants in the full-support group were not simply "doing more practice" but were actually "changing how they approached diagnostic reasoning," which explains why they were able to reduce large errors in ambiguous or borderline cases. This pattern aligns with our design intent to scaffold teachers' causal mental models: XIA makes the evidence—interpretation—action links explicit and testable through contrastive views and counterfactual exploration, enabling a build—test—revise cycle during assessment. Decision-support group, in contrast, likely benefited from additional cues or KCs, which reduced some common mistakes but could not reliably suppress high-risk misjudgments. Control group's lack of systematic support meant that prior strategies and individual heuristics continued to dominate, leaving error patterns unstable.

**Possible psychological mechanisms:** These differences suggest that the observed improvements may reflect a consolidation or shift in teachers' diagnostic mental models, rather than mere effects of exposure or additional practice. We tentatively interpret them through three interrelated mechanisms:

- **Providing causal cues for reflection [72, 88].** Explanations decomposed judgments into operationalizable criteria (e.g., missing prerequisite knowledge vs. question flaw), helping participants transfer these cues to novel cases and reduce the probability of large deviations.
- **Triggering cognitive conflict for recalibration [53, 73, 74].** Contrastive explanation ("why wrong / how to revise") often challenged teachers' initial intuitions (e.g., FSG-04: *"I thought the student had not reviewed, but the explanation showed it might be the question design"*), prompting error monitoring and self-calibration—an effect also captured in the self-regulation scale.
- **Reducing cognitive load [20, 34, 126].** By translating abstract rules into visual reasoning paths and counterfactual examples, the tool lowered working-memory demands in complex scenarios, thereby shrinking variability in error distributions (narrower interquartile ranges, lower tails).

Consistent with this mechanistic view, RMSE decreased more than MAE in full-support group. Because RMSE is particularly sensitive to large deviations, this indicates that explanatory scaffolding helped participants not only get closer on average but also avoid severe misjudgments in borderline cases. Such improvements reflect a deeper shift in teachers' reasoning strategies, aligning with their self-reported gains in reflection and self-regulation [67, 68].

**Divergent user experiences across explanation types:** On the other hand, our in-depth interviews revealed that different explanatory mechanisms evoked distinct experiences among participants. For the contrastive explanation, participants described it as an intuitive way to explore how diagnostic outcomes change. They used this feature to examine which specific items or pieces of evidence were decisive for the model's judgment—why a small change could overturn a diagnosis—and to reflect on item difficulty and evidence weighting (*"I didn't read the question carefully. At first, I thought the student just got lucky, but after flipping the item I realized this question actually mattered,"* FSG-01; *"I was curious why the mastery rate remained 70% even when all answers were correct, and I wanted to understand the model's logic,"* FSG-03). The counterfactual explanation was viewed as a way to test disagreement. Several participants said they used it to check whether their alternative judgment was logically consistent with the data (*"I thought the student's mastery was 30%, but the system showed that this would mean the student got everything wrong, which made me question my own evaluation,"* FSG-02). Others noted that this feature prompted them to consider whether observed errors stemmed from conceptual misunderstanding or mere carelessness (*"It made me consciously think that not all wrong answers mean misunderstanding,"* FSG-04). Although both mechanisms were generally perceived as useful, participants also pointed out their respective limitations. Some found the contrastive function visually appealing and engaging but cognitively demanding; they needed time to understand why full correctness did not necessarily yield 100% mastery, or why small flips sometimes caused large changes (*"It's interesting, but it's a bit hard to grasp why the score sometimes shifts so much,"* FSG-02). A few suggested clearer on-screen prompts or simplified graphics to lower the entry barrier for novice users. In contrast, the counterfactual function was praised for supporting reflection but was also regarded as abstract in form. Some pre-service teachers initially misunderstood it as changing the model's conclusion rather than hypothesizing an alternative causal state (*"I wasn't sure what actually changed when I entered my own answer,"* FSG-04). Others noted that reverse-simulation explanations required a good grasp of diagnostic reasoning, which could be overwhelming for first-time users.

## 8 Limitations and Future Work

While this study provides evidence for the potential of explainable cognitive diagnostic tools in developing pre-service teachers' AL, several limitations remain.

First, the data were collected from a single-session experimental design. The observed improvements therefore primarily reflect short-term intervention effects rather than sustained growth. Future work should adopt longitudinal designs to systematically trace teachers' development across stages and examine the long-term impact of explainable tools on their professional practice in authentic classroom contexts [65].

Second, the study focused on a single subject domain and a specific population of pre-service teachers. It remains unclear whether the findings generalize to teachers in other disciplines or at different levels of teaching experience. Future research could expand to more diverse cohorts—for example, comparing teachers in STEM and humanities subjects, or pre-service and in-service teachers—to reveal both the generalizability and the differentiated impact of explainable tools across contexts.

Third, our analysis primarily relied on response data and system interaction data to capture teachers' assessment processes, which may not fully represent their behavior in real classroom settings. Future studies could integrate multimodal data sources[26], such as classroom video [112], spoken interactions [64], and response records [58], to more comprehensively capture teachers' reasoning pathways and assessment practices. Such evidence would provide a stronger basis for refining tool design and informing teacher training.

Fourth, although the study aimed to examine the role of explanatory scaffolding in teachers' diagnostic reasoning, the current design does not fully isolate the effects of explanatory scaffolding from other intertwined factors, including structured exposure to the assessment tool, repeated assessment practice, and implicit instruction in IRT-based diagnostic concepts. Accordingly, the observed gains in assessment literacy should be interpreted as the outcome of a combined instructional experience rather than the effect of explanation alone. While such integration reflects authentic teacher education settings, it also limits fine-grained causal attribution. In addition, the intervention bundled two explanatory mechanisms—contrastive explanations and counterfactual explanations—making it difficult to disentangle their independent or interactive contributions to the observed gains in assessment literacy. Future work should adopt dismantling or factorial designs, with matched time-on-task and researcher attention, to separately estimate the effects of explanatory scaffolding, structured tool exposure, and assessment practice, as well as the main and interaction effects of different explanation types. Complementary process measures (e.g., fine-grained interaction logs or think-aloud protocols) could further validate mechanism use and clarify underlying causal pathways.

Finally, building on recent advances in XAI, future designs should move beyond static outputs toward domain-driven [43, 85], context-sensitive [101] explanation pipelines: translate model rationales into pedagogically familiar language; embed adaptive, interactive counterfactuals that (when model confidence is high) support cognitive-load reduction; offer switchable explanation styles personalized to teacher characteristics (e.g., need for cognition) [28, 84]; incorporate human-in-the-loop critique to refine and align explanations with teachers' reasoning [103, 145]; and stage delivery in a three-step flow (causal connection → explanation selection → presentation). Such dynamic, personalized, and theory-grounded mechanisms are likely to strengthen understanding and trust, improve judgment calibration, and promote transfer into classroom practice.

## 9 Conclusion

This study addresses a persistent challenge in teacher education—how to effectively cultivate assessment literacy among pre-service teachers—and proposes both design principles and an integrated system-level solution. Drawing on formative interviews and prior theoretical work, we identify core requirements for teacher-facing intelligent assessment tools: providing decision support that is tightly coupled with authentic diagnostic tasks, and making the underlying chains of evidence transparent and traceable so that reasoning processes become observable, examinable, and calibratable. Guided by these principles, we developed XIA, an intelligent assessment platform that integrates multidimensional learning evidence with explainable diagnostic mechanisms to scaffold teachers' professional judgment.

Our mixed-method evaluation yields preliminary insights suggesting that explanatory scaffolding, when embedded in structured assessment tools and practice-oriented tasks, may support growth in several key dimensions of assessment literacy, including reflection, self-regulation, and assessment awareness, and may help mitigate some forms of serious judgment errors. Notably, participants demonstrated a shift from score-oriented judgments toward more evidence-based diagnostic reasoning, indicating that explainable mechanisms can to some extent assist teachers in constructing and calibrating the causal mental models underlying professional judgment.

From a design perspective, this work illustrates how explainability can be embedded into teachers' assessment workflows and outlines a pathway through which explainable AI may contribute to the development of sustainable assessment literacy. XIA offers a technological approach that combines explainable intelligent diagnosis with task-embedded learning to strengthen teachers' diagnostic insight and instructional decision-making. Future research may extend this approach across disciplines, grade levels, and longer-term classroom contexts to further examine how explainable assessment systems can support teachers' ongoing professional growth in authentic educational environments.

## Acknowledgments

We thank the reviewers for their insightful comments and feedback, which helped us to significantly improve the work. This work was supported by the National Natural Science Foundation of China (Grant No. 62477012), the Natural Science Foundation of Shanghai, China (Grant No. 23ZR1418500), the AI for Science Program of the Shanghai Municipal Commission of Economy and Informatization, China (Grant No. 2025-GZL-RGZN-BTBX-01014), and the National Research Foundation, Singapore and Infocomm Media Development Authority (Grant No. DTC-RGC-09).

## A  Formative Research Information

Information about the teachers interviewed in the formative research study is shown in Table 2.

Table 2: In-service Teacher Sample

| In-service Teacher ID | Subject | Teaching Experience (years) | School Type |
| --- | --- | --- | --- |
| InS-T 1 | Mathematics | 30 | Regular junior high school |
| InS-T 2 | Mathematics | 12 | Regular junior high school |
| InS-T 3 | Mathematics | 15 | Regular junior high school |
| InS-T 4 | Mathematics | 32 | Key senior high school |
| InS-T 5 | Mathematics | 26 | Regular senior high school |
| InS-T 6 | Technology | 3 | Regular senior high school |
| InS-T 7 | Technology | 3 | Key senior high school |

## B  XIA Platform Detailed Interface and Functions

### B.1  Detailed Interface and Functions

- A. Click the **overview** button will expand the A1—A4 interfaces.
- A1. Radar chart of knowledge mastery for comparing relative proficiency across knowledge components.
- A2. Item-level accuracy distribution highlighting common errors.
- A3. Weight of each knowledge component in the test, supporting "importance × performance" evaluation.
- A4. Class-level summary with improvement suggestions.
- B. Click the **question** box will expand the B1—B3 interfaces.
- B1. Item statistics (accuracy, difficulty, discrimination) to locate problematic questions.
- B2. Student response distribution revealing error patterns.
- B3. Error analysis with pedagogical suggestions explaining possible conceptual misconceptions.
- C. Click **knowledge** box will expand the C1—C2 interfaces.
- C1. Distribution of related questions by difficulty and mastery.
- C2. Performance analysis with recommended teaching strategies. Together, these views help teachers identify instructional priorities and design class/individual interventions.
- D. Textual diagnostic conclusions with cited evidence.
- E. Visualized reasoning chains from student responses to latent knowledge states
- E1. Clicking a **question** box enables *contrastive explanations* ("if the student had answered differently, what assessment would result?").
- F. Reset to the student's original response state.
- G. Knowledge mastery values.
- H. Disagreement input: teachers can override system predictions and input their own evaluation.
- H1. Enter your assumptions about the student's level of mastery of this knowledge. Click the **Apply** button will expand components H2—H3.
- H2. Counterfactual explanations: showing how predictions and evidence would change if mastery assumptions were altered ("if the student's mastery of this concept were $x$, which questions would they likely succeed or fail on?"). These functions help teachers interpret and critically examine AI assessment outputs, providing a reflective and communicable basis.
- H3. The model infers student performance based on the teacher's opinions.

### B.2  Computational Methods for Diagnosis and Explanation

*Notation.* Let $N$ students, $M$ items, and $K$ knowledge components (KCs). Student and item IDs are one-hot vectors $\mathbf{x}_s \in \{0,1\}^{1 \times N}$ and $\mathbf{x}_e \in \{0,1\}^{1 \times M}$. The Q-matrix $Q \in \{0,1\}^{M \times K}$ encodes item–KC relevancy. We denote the sigmoid by $\sigma(\cdot)$ and the element-wise (Hadamard) product by $\odot$.

Base model (NeuralCDM) — diagnosis engine. *Student and item factors.*

$$\mathbf{h}^s = \sigma(\mathbf{x}_s A) \in (0,1)^{1 \times K}, \quad \mathbf{Q}_e = \mathbf{x}_e Q \in \{0,1\}^{1 \times K},$$

$$\mathbf{h}_e^{\text{diff}} = \sigma(\mathbf{x}_e B) \in (0,1)^{1 \times K}, \quad h_e^{\text{disc}} = \sigma(\mathbf{x}_e D) \in (0,1).$$

Here, $A \in \mathbb{R}^{N \times K}$ (learnable) maps student IDs to pre-activations whose sigmoid gives the mastery vector $\mathbf{h}^s$. $Q \in \{0,1\}^{M \times K}$ is *given and fixed*, so $\mathbf{Q}_e$ is the (non-trainable) KC requirement of item $e$. $B \in \mathbb{R}^{M \times K}$ (learnable) yields KC-wise difficulty $\mathbf{h}_e^{\text{diff}}$ for item $e$, and $D \in \mathbb{R}^{M \times 1}$ (learnable) yields a scalar discrimination gate $h_e^{\text{disc}}$; a KC-wise gate is possible by letting $D$ be $M \times K$ and replacing scalar broadcast with Hadamard product.

*Interaction and prediction (KC-wise).*

$$\mathbf{x} = \mathbf{Q}_e \odot (\mathbf{h}^s - \mathbf{h}_e^{\text{diff}}) \odot h_e^{\text{disc}} \in \mathbb{R}^{1 \times K},$$

$$\mathbf{g}_1 = \sigma(W_1 \mathbf{x}^\top + b_1), \quad \mathbf{g}_2 = \sigma(W_2 \mathbf{g}_1 + b_2), \quad y_{s,e} = \sigma(W_3 \mathbf{g}_2 + b_3).$$

Here, $\mathbf{x}$ is the KC-wise interaction for item $e$; if $h_e^{\text{disc}}$ is scalar, it acts as a positive gate broadcast to each KC. $W_1 \in \mathbb{R}^{H_1 \times K}$, $W_2 \in \mathbb{R}^{H_2 \times H_1}$, $W_3 \in \mathbb{R}^{1 \times H_2}$ and $b_1 \in \mathbb{R}^{H_1}$, $b_2 \in \mathbb{R}^{H_2}$, $b_3 \in \mathbb{R}$ are learnable. To preserve coordinate-wise monotonicity of the logit w.r.t. $\mathbf{h}^s$ on the support of $\mathbf{Q}_e$, we constrain $W_1, W_2, W_3$ to be element-wise nonnegative (e.g., via per-step clipping $W_i \leftarrow \max(W_i, 0)$ or reparametrization $W_i = \text{softplus}(V_i)$); activations use $\sigma(\cdot)$.

*Training objective.*

$$\mathcal{L}_{\text{CE}} = -\sum_{(s,e) \in \Omega} \left[ r_{s,e} \log y_{s,e} + (1 - r_{s,e}) \log(1 - y_{s,e}) \right].$$

Here, $\Omega$ is the set of observed student-item pairs, $r_{s,e} \in \{0, 1\}$ is correctness, and $y_{s,e} \in (0, 1)$ is the predicted correctness probability.

**(a) Test-time posterior update for a new student $T$.** We **freeze** $(Q, B, D, W_1, W_2, W_3, b_1, b_2, b_3)$ and all other students' representations, and **only** estimate $T$'s mastery. To avoid box projections, we use a logistic reparameterization

$$\mathbf{h}_T^s = \sigma(\mathbf{u}_T), \qquad \mathbf{u}_T \in \mathbb{R}^{1 \times K},$$

and minimize the student-specific cross-entropy

$$\widehat{\mathbf{u}}_T = \arg\min_{\mathbf{u} \in \mathbb{R}^{1 \times K}} \sum_{(e,r) \in \mathcal{R}_T} \left[ -r \log y(\sigma(\mathbf{u}), e) - (1 - r) \log(1 - y(\sigma(\mathbf{u}), e)) \right],$$

where $\mathcal{R}_T = \{(e, r)\}$ are $T$'s responses and $y(\cdot, e)$ denotes the model's prediction for item $e$ given a candidate $\mathbf{h}_T^s$. We update $\mathbf{u}_T$ by gradient steps (unconstrained); the posterior mastery is $\widehat{\mathbf{h}}_T^s = \sigma(\widehat{\mathbf{u}}_T)$. (Equivalently, one may add a new row $A_{T,:}$ and only update this row with $\mathbf{h}_T^s = \sigma(\mathbf{x}_T A)$.)

**(b) Contrastive reasoning (same student $T$).** **Goal:** show how different evidence subsets for the *same* student change the diagnosis. Construct two subsets $\mathcal{R}_T^{(1)}$ and $\mathcal{R}_T^{(2)}$. With the same initialization $\mathbf{u}^{(0)}$ and all non-student parameters frozen as in (a), fit

$$\widehat{\mathbf{u}}_T^{(j)} = \arg\min_{\mathbf{u}} \sum_{(e,r) \in \mathcal{R}_T^{(j)}} \left[ -r \log y(\sigma(\mathbf{u}), e) - (1 - r) \log(1 - y(\sigma(\mathbf{u}), e)) \right], \quad j \in \{1, 2\},$$

and set $\widehat{\mathbf{h}}_T^s(\mathcal{R}_T^{(j)}) = \sigma(\widehat{\mathbf{u}}_T^{(j)})$. We report the KC-wise difference

$$\Delta \mathbf{h}_T^s = \widehat{\mathbf{h}}_T^s(\mathcal{R}_T^{(2)}) - \widehat{\mathbf{h}}_T^s(\mathcal{R}_T^{(1)}),$$

which indicates how the estimated mastery changes across the two evidence sets (we do not further attribute the difference to individual items).

**(c) Counterfactual reasoning. Goal:** simulate "what-if" outcomes by *replacing* $T$'s mastery with a counterfactual vector and forwarding through the *fixed* model. Choose a KC subset $S \subseteq \{1, \ldots, K\}$ and define $\mathbf{m}^{\text{cf}} \in (0, 1)^{1 \times K}$ by overwriting entries on $S$ (e.g., grid $\{0.1, 0.3, 0.5, 0.7, 0.9\}$) while keeping others equal to $\widehat{\mathbf{h}}_T^s$. We map to the student embedding with an identity mapping,

$$\mathbf{h}_T^{s*} = \phi(\mathbf{m}^{\text{cf}}), \qquad \phi = \text{Id},$$

and freeze $\Theta_{\text{fixed}} = \{Q, \mathbf{h}_e^{\text{diff}}, h_e^{\text{disc}}, W_1, W_2, W_3, b_1, b_2, b_3, \mathbf{h}_{s \neq T}^s\}$. We then replace $\widehat{\mathbf{h}}_T^s$ by $\mathbf{h}_T^{s*}$ and *only* perform a forward pass per item $e$ (no gradient is taken):

$$\mathbf{x}^{\text{cf}} = \mathbf{Q}_e \odot (\mathbf{h}_T^{s*} - \mathbf{h}_e^{\text{diff}}) \odot h_e^{\text{disc}}, \qquad y'_{T,e} = \sigma\Big(W_3 \sigma\big(W_2 \sigma(W_1 \mathbf{x}^{\text{cf}\top} + b_1) + b_2\big) + b_3\Big).$$

This yields counterfactual per-item correctness probabilities $\{y'_{T,e}\}$; if a binary pattern is required, we threshold the probabilities (e.g., at 0.5 or a calibrated cutoff).

## C Detailed Experimental Data on Assessment Accuracy and Questionnaires

### C.1 Detailed Data

*C.1.1 Participant Statistics.* All participants were pre-service high-school teachers specializing in Mathematics and Technology Education. None had taken on full-time teaching positions; their teaching experience came primarily from school-based practicum, with an average internship duration of approximately half a year (see Table 3). Regarding familiarity with assessment concepts, all participants achieved full scores on the scenario-based assessment test in the questionnaire (Appendix C.2 Part 1), indicating that they possessed basic knowledge related to educational assessment. Participants voluntarily enrolled in the study through both online and offline recruitment announcements. As pre-service teachers who were about to graduate and transition into full-time teaching roles, they expressed a strong interest in using the study as an opportunity to practice interpreting student response data, experience digital assessment tools, and enhance their assessment literacy and instructional decision-making skills in realistic task contexts. No course credit or additional academic incentives were offered, and participation was entirely voluntary.

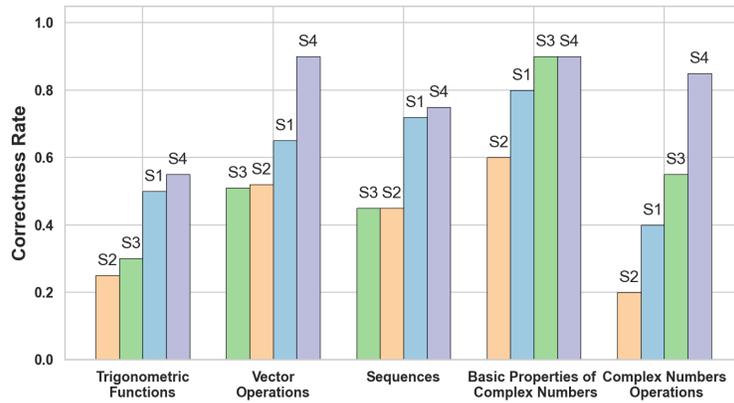

Figure 8: KC-level correctness rates ordered by students' mastery estimates. X-axis: knowledge components; Y-axis: correctness rate for each student (S1—S4), with bars ordered by estimated mastery within each KC.

Table 3: Participant demographics across conditions (computed from the pre-study questionnaire).

| Group | N | Background | Age (M ± SD) | Internship (years M ± SD) |
|---|---|---|---|---|
| Control Group | 7 | Math/Tech Education (100.0%) | 25.0 ± 2.9 | 0.50 ± 0.40 |
| Decision-Support Group | 7 | Math/Tech Education (100.0%) | 22.6 ± 1.6 | 0.50 ± 0.40 |
| Full-Support Group | 7 | Math/Tech Education (100.0%) | 24.3 ± 2.7 | 0.54 ± 0.28 |

*Note.* "Background" summarizes whether the major is related to mathematics/technology education; the percentage indicates the share within each group. "Internship" is the original internship duration from the questionnaire. Values are the means and standard deviations.

*C.1.2 Baseline for Students' Knowledge Mastery.* Fig. 8 illustrates the trend consistency between model-estimated mastery levels and students' KC-level learning outcomes. For each KC, the four participating students (S1—S4) are ordered from left to right based on their estimated mastery for that KC. The height of each bar shows the corresponding correctness rate derived from their item-level responses. Across all five KCs, students with higher mastery estimates consistently demonstrate higher correctness rates, providing behavioral evidence that the diagnostic reference aligns with observable performance patterns. This supports the use of the mastery profile as a quasi–ground truth for evaluating pre-service teachers' diagnostic judgments in the main study.

*C.1.3 Analyze Statistics.* To complement the main analyses reported in the paper, this appendix provides detailed within- and between-group statistics for each subscale, along with results from both parametric ($t$-tests) and non-parametric tests (Wilcoxon, Mann-Whitney U). The findings show consistent significance outcomes across both approaches, offering dual support for the robustness of the conclusions.

Table 4: Shapiro—Wilk normality tests on gain scores (Δ = Post—Pre).

| Measure | Subscale | W | p | Normal ($p > 0.05$) |
|---|---|---|---|---|
| Accuracy | MAE | 0.968 | 0.679 | Yes |
|  | RMSE | 0.972 | 0.780 | Yes |
| Questionnaire | Reflection | 0.927 | 0.121 | Yes |
|  | Self-regulation | 0.962 | 0.552 | Yes |
|  | Assessment awareness | 0.908 | 0.053 | Yes |

*Note.* Shapiro-Wilk tests were performed on gain scores (Δ = Post—Pre) to examine the normality assumption for $t$-tests. Group-level tests ($n = 7$) showed occasional deviations, but robustness checks using Wilcoxon/MWU yielded consistent conclusions.

**Within-group comparisons of assessment accuracy (Table 5).** Full-support group showed the most consistent pre—post reductions across both error metrics. For MAE, the decrease was substantial and statistically significant ($\Delta M = -0.063$, $t = 3.78$, $p = 0.009$), and for

Table 5: Within-group pre—post comparisons of assessment accuracy: *t* test and Wilcoxon signed-rank test.

| Metric | Group | Pre M | Post M | Mean Diff | t | p (t) | p (Wilcoxon) |
|---|---|---|---|---|---|---|---|
| MAE | CG | 0.148 | 0.135 | -0.013 | 0.64 | 0.547 | 0.813 |
| | DSG | 0.176 | 0.143 | -0.033 | 0.80 | 0.453 | 0.578 |
| | FSG | 0.165 | 0.102 | -0.063 | 3.78 | 0.009 | 0.016 |
| RMSE | CG | 0.180 | 0.159 | -0.022 | 1.06 | 0.329 | 0.578 |
| | DSG | 0.204 | 0.170 | -0.034 | 0.77 | 0.472 | 0.578 |
| | FSG | 0.190 | 0.124 | -0.067 | 3.04 | 0.023 | 0.047 |

*Note.* Negative mean differences indicate error reduction (post < pre). Significant improvements ($p < 0.05$) appear in the full-support group. CG: Control Group; DSG: Decision-Support Group; FSG: Full-Support Group.

RMSE, the reduction was also substantial ($\Delta M = -0.067$, $t = 3.04$, $p = 0.023$). Decision-support group exhibited smaller, non-significant decreases in both MAE ($\Delta M = -0.033$, $t = 0.80$, $p = 0.453$) and RMSE ($\Delta M = -0.034$, $t = 0.77$, $p = 0.472$). Control group showed little to no improvement (MAE: $\Delta M = -0.013$, $t = 0.64$, $p = 0.547$; RMSE: $\Delta M = -0.022$, $t = 1.06$, $p = 0.329$). Thus, only the full-support group achieved statistically reliable and practically meaningful reductions in assessment error.

Table 6: Between-group comparisons of assessment accuracy gain scores (Δ = post—pre): Welch *t* test and Mann—Whitney U test.

| Metric | Contrast(g1 vs g2) | Δ M (g1) | Δ M (g2) | Diff | Welch t | p (t) | p (MWU) |
|---|---|---|---|---|---|---|---|
| MAE | DSG vs FSG | -0.033 | -0.063 | 0.030 | 1.75 | 0.117 | 0.179 |
| | DSG vs CG | -0.033 | -0.013 | -0.020 | 0.35 | 0.736 | 1.000 |
| | FSG vs CG | -0.063 | -0.013 | -0.050 | 0.77 | 0.462 | 0.165 |
| RMSE | DSG vs FSG | -0.034 | -0.067 | 0.033 | 1.64 | 0.142 | 0.209 |
| | DSG vs CG | -0.034 | -0.022 | -0.012 | 0.29 | 0.780 | 0.754 |
| | FSG vs CG | -0.067 | -0.022 | -0.045 | 1.03 | 0.323 | 0.209 |

*Note.* Negative Δ values reflect reductions in error. No between-group contrasts reached significance at $p < 0.05$. CG: Control Group; DSG: Decision-Support Group; FSG: Full-Support Group.

**Between-group comparisons of assessment accuracy gain scores (Table 6).** Analyses of gain scores suggested that the full-support group tended to outperform the control group. For MAE, the full-support group's advantage over the control group was moderate but not significant (ΔDiff = −0.050, Welch $t = 0.77$, $p = 0.462$). For RMSE, full-support group again exceeded the control group by ΔDiff = −0.045, though the difference did not reach significance ($t = 1.03$, $p = 0.323$). Comparisons between the full-support group and the decision-support group showed non-significant trends favoring the full-support group (MAE: ΔDiff = −0.030, $t = 1.75$, $p = 0.117$; RMSE: ΔDiff = −0.033, $t = 1.64$, $p = 0.142$). Decision-support group did not significantly outperform the control group (MAE: ΔDiff = −0.020, $p = 0.736$; RMSE: ΔDiff = −0.012, $p = 0.780$). Overall, the pattern of results suggests that full-support group consistently showed the largest error reductions, but the small sample size limited between-group significance.

**Within-group comparisons of the questionnaire (Table 7).** Across the three subscales, the full-support group (FSG) showed the most consistent pre—post gains: Reflection increased by $\Delta M = 0.86$ ($t = 9.30$, $p < 0.001$), Self-regulation by $\Delta M = 0.93$ ($t = 4.20$, $p = 0.0057$), and Assessment awareness by $\Delta M = 0.68$ ($t = 9.46$, $p < 0.001$). The decision-support group (DSG) also improved in Self-regulation ($\Delta M = 0.81$, $t = 4.72$, $p = 0.0032$) and Assessment awareness ($\Delta M = 0.34$, $t = 3.44$, $p = 0.0138$), with a smaller gain in Reflection ($\Delta M = 0.70$, $t = 2.58$, $p = 0.0417$). The control group (CG) showed little change in Reflection ($\Delta M = 0.04$, $t = 0.19$, $p = 0.8528$) and Self-regulation ($\Delta M = 0.13$, $t = 0.79$, $p = 0.4600$); the increase in Assessment awareness did not reach significance ($\Delta M = 0.25$, $t = 1.89$, $p = 0.1082$). Wilcoxon signed-rank tests converged on the same conclusions (Table 7). We intentionally avoid interpreting paired standardized effect sizes that divide by the SD of the difference, because in this design (same items, same session, self-reports) pre-post responses can be highly correlated and/or the pre/post SDs can be small, yielding very small $\Delta SD$ (e.g., 0.19-0.59 in FSG) and mechanically inflating such standardized indices; instead, we focus on absolute change and its scale-bound meaning via the Percentage of Maximum Possible (POMP) [25]. On a 1—7 scale (range = 6), the FSG gains of 0.68-0.93 correspond to 11—16% of the maximum possible improvement (last column), a practically meaningful change that is consistent with the only-moderate improvements observed on the objective accuracy metrics (MAE/RMSE). Treating summated Likert scores as approximately interval for these analyses follows standard practice [100].

**Between-group comparisons of questionnaire gain scores (Table 8).** Analyses of gain scores suggested that the full-support group (FSG) consistently outperformed the control group (CG). For Reflection, the full-support group's advantage over the control group was substantial and statistically significant ($\Delta M = 0.82$, Welch $t = 3.98$, $p = 0.0033$). For Self-regulation, the full-support group again surpassed the control group ($\Delta M = 0.80$, Welch $t = 2.91$, $p = 0.0141$). Similarly, for Assessment awareness, the full-support group's gains exceeded

Table 7: Within-group questionnaire comparisons (Δ = post−pre): $t$ test and Wilcoxon signed-rank test.

| Subscale | Group | Δ M | ΔSD | $r$ | $t$ | $p$ (t) | $p$ (Wilcoxon) | POMPΔ% |
|---|---|---|---|---|---|---|---|---|
| Reflection | CG | 0.04 | 0.49 | 0.79 | 0.19 | 0.8528 | 1.0000 | 0.6 |
|  | DSG | 0.70 | 0.71 | -0.07 | 2.58 | 0.0417 | 0.0312 | 11.6 |
|  | FSG | 0.86 | 0.24 | -0.22 | 9.30 | 0.0001 | 0.0156 | 14.3 |
| Self-regulation | CG | 0.13 | 0.43 | 0.27 | 0.79 | 0.4600 | 0.5992 | 2.1 |
|  | DSG | 0.81 | 0.46 | 0.77 | 4.72 | 0.0032 | 0.0156 | 13.6 |
|  | FSG | 0.93 | 0.59 | 0.89 | 4.20 | 0.0057 | 0.0156 | 15.5 |
| Assessment awareness | CG | 0.25 | 0.35 | 0.81 | 1.89 | 0.1082 | 0.0679 | 4.2 |
|  | DSG | 0.34 | 0.26 | 0.78 | 3.44 | 0.0138 | 0.0277 | 5.7 |
|  | FSG | 0.68 | 0.19 | 0.73 | 9.46 | 0.0001 | 0.0156 | 11.4 |

*Note.* All significant results ($p < 0.05$) indicate improvements from pre- to post-test. Results from $t$ tests and Wilcoxon signed-rank tests converge on the same conclusions. $r$ denotes the within-pair correlation between pre- and post-test scores for each group. CG: Control Group; DSG: Decision-Support Group; FSG: Full-Support Group; POMP: Percentage of Maximum Possible.

Table 8: Between-group comparisons of questionnaire gain scores (Δ = post−pre): Welch $t$ test and Mann–Whitney U test.

| Subscale | Contrast (g1 vs g2) | ΔM | ΔSD | $r_{g1}$ | $r_{g2}$ | Welch $t$ | $p$ (t) | $p$ (MWU) | POMPΔ%(g1-g2) |
|---|---|---|---|---|---|---|---|---|---|
| Reflection | FSG vs DSG | 0.16 | 0.53 | -0.21 | -0.07 | 0.56 | 0.5897 | 0.0890 | 2.7 |
|  | FSG vs CG | 0.82 | 0.39 | -0.21 | 0.79 | 3.98 | 0.0033 | 0.0142 | 13.7 |
|  | DSG vs CG | 0.66 | 0.61 | -0.07 | 0.79 | 2.02 | 0.0692 | 0.0825 | 11.0 |
| Self-regulation | FSG vs DSG | 0.11 | 0.53 | 0.89 | 0.77 | 0.41 | 0.6912 | 1.0000 | 1.9 |
|  | FSG vs CG | 0.80 | 0.51 | 0.89 | 0.27 | 2.91 | 0.0141 | 0.0289 | 13.3 |
|  | DSG vs CG | 0.69 | 0.44 | 0.77 | 0.27 | 2.89 | 0.0136 | 0.0250 | 11.4 |
| Assessment awareness | FSG vs DSG | 0.34 | 0.23 | 0.73 | 0.78 | 2.78 | 0.0180 | 0.0250 | 5.7 |
|  | FSG vs CG | 0.43 | 0.29 | 0.73 | 0.81 | 2.82 | 0.0197 | 0.0398 | 7.1 |
|  | DSG vs CG | 0.09 | 0.31 | 0.78 | 0.81 | 0.53 | 0.6083 | 0.4763 | 1.5 |

*Note.* Significant differences ($p < 0.05$) appear mainly when comparing the full-support group to the control group. Mann–Whitney U tests replicate these results, supporting the robustness of the findings. CG: Control Group; DSG: Decision-Support Group; FSG: Full-Support Group; Δ M: The difference in mean pre−post change between the two groups; Δ SD: The pooled standard deviation of the Δ for g1 and g2, summarizing the common dispersion of their deltas.

the control group's by 0.43 (Welch $t$ = 2.82, $p$ = 0.0197). Comparisons between the full-support group and the decision-support group (DSG) showed smaller differences, reaching significance only in Assessment awareness (ΔDiff = 0.34, Welch $t$ = 2.78, $p$ = 0.0180). The decision-support group also outperformed the control group in Self-regulation (ΔDiff = 0.69, Welch $t$ = 2.89, $p$ = 0.0136), but not in Reflection ($p$ = 0.0692) or Assessment awareness ($p$ = 0.6083). Consistent with recommendations for same-session pre–post designs with highly correlated self-reports, the same considerations apply to between-group gain-score comparisons; therefore, we do not interpret standardized effect sizes and instead focus on absolute mean differences and POMP for scale-relative interpretation.

**Robustness of results.** Across both within- and between-group analyses, the non-parametric tests yielded conclusions consistent with the parametric tests. For the within-group comparisons, full-support group's reductions in both MAE and RMSE were significant by paired $t$ tests ($p$ = 0.009 and $p$ = 0.023) and corroborated by Wilcoxon signed-rank tests ($p$ = 0.016 and $p$ = 0.047). In contrast, the decision-support group and the control group showed no significant changes under either method. For the between-group contrasts, the same pattern was observed: for example, on Reflection, the full-support group vs. the control group was significant both by Welch $t$ ($p$ = 0.0033) and Mann–Whitney U ($p$ = 0.0142) in a parallel analysis, and similar convergence appeared for Self-regulation (full-support group vs. control group; $p$ = 0.0141 vs. $p$ = 0.0289) and Assessment awareness (full-support group vs. decision-support group; $p$ = 0.0180 vs. $p$ = 0.0250). This concordance across test types indicates that the observed effects are robust to distributional assumptions. Moreover, the magnitudes of change for significant contrasts were consistently non-trivial (POMPΔ% > 10% within groups, POMPΔ% > 7% between groups). These magnitudes provide preliminary evidence of the tool's practical significance.

### C.2 Questionnaire Content

We employed questionnaire instruments adapted from validated scales with established reliability and validity, making minimal contextual modifications to fit the XIA scenario. In the present sample, we re-examined internal consistency for the three multi-item subscales (see Table 9). Cronbach's $\alpha$ values were 0.641, 0.720, and 0.573, respectively. Because Cronbach's $\alpha$ is sensitive to the assumptions of interval-level measurement and $\tau$-equivalence, and tends to underestimate reliability for ordinal Likert-type data, we additionally reported ordinal $\alpha$ based on Spearman correlations, which were 0.661, 0.814, and 0.712, respectively. Although the third subscale showed a relatively low Cronbach's $\alpha$ of 0.573, its ordinal $\alpha$ of 0.712 met the conventional acceptability threshold, indicating adequate internal consistency across subscales in this sample. The study measured four dimensions:

Table 9: Internal consistency of the three subscales.

| Subscale | Cronbach's $\alpha$ | Ordinal $\alpha$ (Spearman) |
| --- | --- | --- |
| Reflection | 0.641 | 0.661 |
| Self-regulation | 0.720 | 0.814 |
| Assessment awareness | 0.573 | 0.712 |

---

**Example Item: Part 1. Instructional scenarios**

**Scenario 1.** Mr. Zhang was grading an exam question on "function zeros" and noticed that many students answered incorrectly. He wanted to determine whether students truly misunderstood the relationship between the function graph and its zeros, or whether they were simply careless in their calculations.
*Question:* Which of the following would be the most effective way to collect additional evidence?

  A. Check whether students' handwriting is neat.
  B. Talk with students to ask if they were serious when solving the problem.
  C. Add a follow-up question asking students to write out their solution steps and explain each step.
  D. Convert the question into multiple choice for easier objective scoring.

**Scenario 2.** On the cognitive diagnosis platform, Mr. Zhang examined Student A's mastery level on the "function zeros" concept, which was only 38%. In Question 8, the student selected distractor option "B," a typical error representing the misconception of misidentifying the axis of symmetry as the zero point.
*Question:* Which of the following instructional judgments is most reasonable?

  A. The student was careless and should practice more similar questions for fluency.
  B. The question is too difficult and should be removed from the question bank.
  C. The student holds a structural misconception on this concept; the teacher should address the distractor and explain the incorrect reasoning.
  D. The student's overall score is acceptable, so this single error can be ignored.

---

**Part 1. Instructional scenarios.** Following the design principles of the Assessment Literacy Inventory (ALI) [93], which uses 5 scenarios × 7 questions aligned with seven dimensions of teacher AL, we adapted the contextualized multiple-choice format. The original ALI classroom scenarios were replaced with authentic tasks in which pre-service teachers read outputs from the cognitive diagnosis platform and made instructional decisions. Questions targeted four dimensions: (a) error interpretation/diagnosis, (b) question discrimination and quality, (c) coverage and alignment with learning objectives, and (d) intervention for individual vs. class differences. These questions directly assessed whether teachers could correctly interpret indicators such as mastery estimates, distractor options, discrimination indices, and KC contributions, and take reasonable actions accordingly.

**Part 2. Teacher reflection scale.** We adopted the four-dimensional structure from Kember's Reflection Questionnaire [69]: habitual action, understanding, reflection, and critical reflection. Item wording and subscales were kept consistent with the original tool, with only the context adapted to "learning and using the assessment/diagnosis tool."

---

**Example Item: Part 2. Teacher reflection scale**

**Response scale:** 1 = Strongly disagree   …   7 = Strongly agree

- As long as I can memorize the key points of course materials, I do not need to think deeply about assessment methods.
- I often reflect on whether my assessment methods can authentically capture students' abilities.

---

**Part 3. Teacher self-regulated learning behaviors.** Drawing on the Self-Regulated Learning Opportunities Questionnaire for Teacher Educators (SRLOQ) [138], which evaluates pre-service teachers' self-regulation opportunities (e.g., planning, monitoring), we refocused the scale on tasks of "assessment design and diagnostic tool use." Items were organized into sub-dimensions: planning, monitoring, time management, task perception, adjustment, reflection, and attribution.

> **Example Item: Part 3. Teacher self-regulated learning behaviors**
> **Response scale:** 1 = Strongly disagree   ...   7 = Strongly agree
> - The learning activities I design are aligned with and serve the achievement of assessment goals.
> - When I realize that my assessment design is not effective, I actively seek alternative strategies.

**Part 4. Teacher assessment awareness.** We adopted the four-factor structure of the Conceptions of Assessment III (CoA-III) [15, 16]: improvement (including sub-factors such as improving learning/teaching, validity/quality, and describing ability), school accountability, student accountability, and irrelevance. The instrument has demonstrated good model fit and internal consistency. We contextualized the items to secondary classrooms and the use of diagnostic tools.

> **Example Item: Part 4. Teacher assessment awareness**
> **Response scale:** 1 = Strongly disagree   ...   7 = Strongly agree
> - I believe that assessment can stimulate students to think and improve their ways of learning.
> - Assessment is merely an accessory step after teaching and does not require much effort to design.

All subscales were administered in pre- and post-tests with isomorphic forms (identical structures with minor contextual word substitutions). Dimension scores were computed as mean values: Part 1 was scored by accuracy, while Parts 2—4 used a 7-point Likert scale.

## D Detailed Interview Information

### D.1 Interview Detailed Analysis

*D.1.1 Common Understandings Across Participants.* Overall, most pre-service teachers reported a shift from a *single-score orientation* toward valuing *multi-dimensional evidence and question quality*. Instead of focusing only on correctness or scores, they considered question difficulty/discrimination, knowledge coverage, and diagnostic validity, and attempted to justify their judgments with data and evidence. Typical comments include (CG: Control Group; DSG: Decision-Support Group; FSG: Full-Support Group.):

> FSG-02: "*Evaluation of a knowledge should take question difficulty and discrimination into account.*"
> DSG-03: "*I used to only look at overall accuracy, but now I also consider the effect of difficulty and discrimination.*"
> CG-01: "*Previously I relied on accuracy rates and intuition; now I pay more attention to question-knowledge mapping.*"

Several participants also began to see assessment as a tool for *promoting learning* (e.g., diagnosing weaknesses) and proposed methodological refinements such as visualization and fine-grained coverage:

> DSG-01: "*My thinking has shifted from single-variable to multi-variable, integrated evaluation.*"
> DSG-07: "*I would use data visualization to check student learning progress.*"
> CG-07: "*I no longer judge only by scores; I also consider whether questions are appropriate and knowledge selection is adequate.*"

**Summary:** The most common consensus was *multi-dimensional + question quality + evidence-based reasoning*, followed by a recognition of *assessment for learning* and methodological attention to visualization/design optimization.

*D.1.2 Differences Across Groups.* Responses from full-support group emphasized evidence calibration and weighting of question attributes. Their language repeatedly referenced difficulty, discrimination, numbers, and evidence, reflecting a shift from intuition to evidence-based judgment:

> FSG-03: "*The numerical values provided by the tool helped me organize my thinking and avoid wavering.*"
> FSG-04: "*I used to rely on my own experience, now I calibrate with the platform information.*"
> FSG-01: "*I no longer just look at correctness rates—I combine question difficulty and discrimination*"
> FSG-02: "*I used to roughly estimate from accuracy, but now I also consider reasoning and question weights.*"

These statements indicate that full-support group more pragmatically internalized evidence-based judgment and question-level diagnosis. They rarely mentioned "scaffolding" explicitly, but described actionable strategies such as weighting and calibration.

Decision-support group also showed cognitive expansion, but mainly at the level of information and presentation. Their answers revealed inspirations (e.g., visualization, noticing question difficulty) but less integration into a stable diagnostic-intervention framework:

> DSG-07: "*Data visualization helps me examine student learning.*"
> DSG-01: "*Some questions are too difficult and may interfere with judgment.*"
> DSG-03: "*I try to consider difficulty and discrimination, not just overall accuracy.*"
> DSG-02: "*I moved from considering only one variable to evaluating through a combination of information.*"

Decision-support group "saw more information," but less often transformed it into robust diagnostic and intervention strategies. Control group participants articulated some awareness of holistic and process-oriented evaluation, but often at an abstract level without operational strategies:

CG-02: "*We need to include process data and student self-reports for a multi-faceted understanding.*"
CG-07: "*I no longer rely solely on scores; I also check indicators and coverage.*"
CG-04: "*Score-only information in the past may have led to misjudgments.*"

Their reflections were conceptual but lacked execution pathways; many responses remained brief or incomplete. Through the above analysis, we can get a summary of cross-group comparison:

- Commonality: all three groups frequently mentioned "multi-dimensional evidence and question quality."
- Full-support group: emphasized evidence calibration and question weighting (*how to do it*); less rhetoric, more actionable strategies.
- Decision-support group: focused on information and presentation (e.g., visualization), but their diagnostic-intervention "loop" was underdeveloped.
- Control group: highlighted comprehensive principles, but lacked executable logic and stable strategies.

*D.1.3 Correspondence with Quantitative Findings.* The qualitative findings aligned with the quantitative results reported in Sections 6.1 and 6.2, though expressed in different forms:

- **Assessment literacy (6.1).** Full-support group did not explicitly use terms such as "scaffolding" or "self-regulation," but their emphasis on evidence calibration, difficulty / discrimination, and weighting reflected stronger reflective and self-regulatory tendencies. Decision-support group recognized multi-dimensionality but mainly at the level of seeing more information. Control group remained at the level of abstract advocacy.
- **Assessment accuracy (6.2).** Full-support group's discourse of "evidence calibration + question weighting" corresponded to improved accuracy (RMSE tail shrinkage and MAE reduction). Decision-support group's "information gain without full closure" matched partial improvements (median reduction but limited variance shrinkage). Control group's "principles without practice" corresponded to no significant improvements.
- **Surface discrepancy.** Although the decision-support group and the control group mentioned "scaffolding and improvement" more often, full-support group operationalized scaffolding into concrete strategies, which translated into stable performance gains and fewer large errors. This represents a gap between *claim vs. execution*: Control group claimed comprehensiveness but lacked implementation, thus showing no quantitative gains; Full-support group articulated fewer slogans but enacted concrete practices, resulting in measurable improvements.

**In short:** Qualitative and quantitative findings were overall convergent. The gap between "what participants say" and "what they do" explains performance differences: Full-support group operationalized explanatory information into evaluative criteria, leading to robust accuracy gains; Decision-support group stopped at the informational level, leading to limited improvement; Control group remained at abstract advocacy without executable strategies, thus showing little improvement.

## D.2 Interview Questions Content

Examples of interview questions are shown in Table 10, and the complete interview questions are in the supplementary materials.

Table 10: Example item: Interview questions, follow-up probes, and coding categories.

| ID | Interview question | Follow-up probes | Coding category |
| --- | --- | --- | --- |
| Decision-support group-(1) / Full-support group-(1) | In your usual student assessment practices, do you actively pay attention to the types of information shown in the *decision-support* interface (e.g., mastery of knowledge, error distribution, typical error types)? Why or why not? | — Were these types of information something you would normally consider without the system, or only because the system presented them?<br>— Can you give an example of when you would or would not pay attention to such information? | Information adoption and trust (pre-existing vs. newly introduced) |
| Full-support group-(5) | Have you ever disagreed with the system's diagnostic results? If so, did you use the *disagreement input* feature? What thoughts did the system's counterfactual explanations trigger for you? | — Would you normally think in such a "counterfactual" way yourself?<br>— How did these explanations help or hinder your understanding of students' reasoning? | Cognitive conflict and reflection |
| Control group-(1) | When assessing students in everyday practice, what kinds of information do you focus on most (e.g., mastery, error types, problem-solving process)? Why? | — Do you prioritize among these types of information?<br>— Are there specific "key cues" that you always check? | Information adoption and trust |